\documentclass[PAPER, atlasdraft=false, USenglish, cernpreprint, texmf, orcidlogo]{atlasdoc}
\usepackage[subcaption=true]{atlaspackage}
\usepackage{atlasbiblatex}

\usepackage{atlasphysics}

\usepackage{float} %
\usepackage{diagbox, eqparbox, hhline}
\usepackage{multirow}
\usepackage{xfrac}
\usepackage{setspace}

\graphicspath{{logos/}{figures/}}

\usepackage{ANA-HION-2024-11-PAPER-defs}

\AtlasTitle{Characterization of nuclear breakup as a function of hard-scattering kinematics using dijets measured by ATLAS in \(p\)+Pb collisions}

\AtlasJournal{\PLB}
\AtlasJournalRef{\PLB 877 (2026) 140440}
\AtlasDOI{10.1016/j.physletb.2026.140440}

\AtlasAbstract{%
This Letter analyzes the sensitivity of event geometry estimators to the initial-state kinematics of hard scattering in proton--lead collisions.
This analysis uses dijets as a proxy for the parton--parton scattering configuration, correlating it with event geometry estimators, namely the energy deposited in the Zero-Degree Calorimeter and the transverse energy recorded in the Forward Calorimeter in the Pb-going direction.
The analysis uses data recorded by the ATLAS detector at the Large Hadron Collider with a nucleon--nucleon center-of-mass energy of 8.16 TeV, corresponding to an integrated luminosity of 56~nb\(^{-1}\).
The jets are measured within the pseudorapidity interval \(-\)2.8 \(<\) \(\eta\) \(<\) 4.5, where positive \(\eta\) values correspond to the direction of the proton beam.
Results are presented as a function of the Bjorken-\(x\) of the parton originating from the proton, \(x_{p}\).
Both event geometry estimators are found to be dependent on \(x_{p}\), with the energy deposited in the Zero-Degree Calorimeter about six times less sensitive to \(x_{p}\) compared with the transverse energy deposited in the Forward Calorimeter.
}

\AtlasRefCode{HION-2024-11}

\PreprintIdNumber{CERN-EP-2025-077}

\hypersetup{pdftitle={ATLAS document},pdfauthor={The ATLAS Collaboration}}

\begin{document}

\maketitle

\tableofcontents

%
%

%
%
%

%
\section{Introduction}
The study of hard processes in proton--lead (\pxPb) collisions at the Large Hadron Collider (LHC)~\cite{Evans:2008zzb} has contributed to a richer understanding of nuclear matter, both in the hot, dense medium produced in nucleus--nucleus collisions and in the cold nuclear environment~\cite{Nagle:2018nvi, Albacete:2017qng}.

The modification of parton distribution functions in the nuclear environment (nPDFs) \cite{Ethier:2020way, Klasen:2023uqj} has been studied through measurements across multiple channels by experiments at the LHC~\cite{CMS-HIN-13-007, HION-2013-09,CMS-HIN-13-001,CMS-HIN-14-001, HION-2022-15-Published,TOPQ-2023-32,CMS-HIN-17-002, LHCb:2017yua} and at the Relativistic Heavy Ion Collider (RHIC)~\cite{PHENIX-NPDF1,PHENIX-NPDDF2,STAR-NPDF1,STAR-NPDF2}.
One important reason to study the modification of PDFs within nuclei is to understand the production rates of hard probes, like jets and heavy vector bosons, in deep inelastic scattering on nuclei and proton--ion ($p$+A) collisions.

Hard process rates have also been studied as a function of \pxPb\ collision centrality, defined using ``global'' event-level variables sensitive to the collision geometry, such as forward transverse energy %
or the number of forward neutrons.
Both of these variables are expected to be sensitive to the overall collision geometry, particularly the number of binary collisions ($N_{\mathrm{coll}}$)
experienced by the projectile proton~\cite{ALICE:2014xsp}.
This assumption was validated by ATLAS in a Z boson measurement~\cite{HION-2013-09}, where the inclusive Z yields were found to be approximately proportional to $N_{\mathrm{coll}}$. Centrality was defined by the event activity, which is the transverse energy measured in the Forward Calorimeter (FCal) facing the nuclear debris~\cite{HION-2012-13}.
However, such linearity in different centrality selections was \textit{not} observed for jets.
ATLAS observed an event activity bias in the first measurement of inclusive jet production in \pxPb\ collisions at 5.02~\TeV ~\cite{ATLAS:2014cpa}.
The analysis reported a significant suppression of the jet production normalized to $N_{\mathrm{coll}}$ in central events compared with peripheral events, with centrality defined using the event activity. The suppression was found to be a function of the jet kinematics.  More specifically, the central-to-peripheral ratio ($R_{\mathrm{CP}}$) was found to be a function of the jet energy. %
Similar observations were made at RHIC by PHENIX~\cite{PHENIX:2015fgy} and, more recently, by STAR~\cite{STAR:2024nwm}.
These results were interpreted as evidence that protons in a configuration containing a parton with a large partonic momentum fraction (\bjork),  when interacting with a nuclear target,  have a significantly smaller than average cross-section and size~\cite{Alvioli:2017wou}, a manifestation of color fluctuation (CF) effects~\cite{Mueller:1982, Brodsky:1982, PhysRevLett.47.297}.
CFs in QCD refer to variations in the transverse area occupied by color fields, which become spatially compact when a single parton carries a large fraction of the hadron’s momentum~\cite{Brodsky:2022bum}.
By causing the proton to interact with fewer nucleons in the nucleus, the CF effects were found to induce a bias on the event activity and, therefore, on the forward transverse energy used to define the centrality. This bias depends on the kinematics of the jets, suggesting that fluctuations in the initial state can weaken the correlation between impact parameter and centrality.
More recently, ATLAS probed this effect in greater detail by analyzing the centrality dependence of dijet production in \pxPb\ data collected at 8.16~\TeV~\cite{ATLAS:2023zfx}.
By leveraging dijets to reconstruct the event-level initial-state kinematics, this analysis directly showed a correlation between the event activity bias and the Bjorken-$x$ of the proton, \xp, consistent with the CF interpretation proposed in Ref.~\cite{Alvioli:2017wou}.
In comparison, the linear scaling of Z boson production with $N_{\mathrm{coll}}$ was validated in a kinematic region characterized by a proton configuration with \xp\ < 0.1, away from where CF effects are most visible~\cite{HION-2013-09}. Furthermore, the Z boson $R_{\mathrm{CP}}$ measured as a function of rapidity shows hints of a correlation with \xp\ in specific higher-rapidity selections, where a significant fraction of the Z bosons also sample large-\xp\ configurations.

The event activity bias, identified in jets, poses significant challenges in identifying jet quenching signatures in small systems like \pxA. Events with identical impact parameter but different proton configurations can end up in different centrality classes.  Since any jet quenching signal in \pxPb\ collisions is small and the event activity bias can be larger, this ambiguity complicates the interpretation of event activity selected \pxPb\ measurements.
Thus, it becomes a topic of great interest to extract the centrality dependence of hard process rates using observables that are less sensitive to initial-state effects~\cite{Perepelitsa:2024eik}.
One alternative approach is to select different centrality classes in \pxA\ events based on the number of spectator neutrons measured at $|\eta|\gtrsim 8$ with a Zero-Degree Calorimeter (ZDC) instrumented in the nucleus-going direction~\cite{PERF-2007-01,Cortese_2019}.
ZDCs have been used as a way to select central events in searches for energy loss signatures, with the measurements in Refs.~\cite{ALICE:2017svf, ATLAS:2022iyq} motivated by the argument that this classification is less biased than the event activity-based selection method.

A measurement of forward neutron production in \pxPb, published by ALICE~\cite{ALICE:2021poe}, reported the energy accumulated in the neutron ZDC as a function of the estimated number of binary collisions.
The results, self-normalized to the average in minimum bias (MB) events in \pxPb\ collisions, found the energy in the Pb-going neutron ZDC to monotonically increase with the number of binary collisions, showing a correlation between these two quantities.
Despite this, the dynamics of the nuclear breakup in \pxPb\ collisions are still not well understood in many aspects, and are poorly modeled by event generators,
especially in the presence of a hard scattering.
Recently, the authors of Ref.~\cite{Alvioli:2024cmd} argued that CF effects may also be relevant in the nuclear breakup process in ultra-peripheral collisions (UPCs) characterized by a resolved photon exchange, linking the number of nucleons wounded by interactions with a vector meson with the nuclear breakup dynamics.
Such effects, related to the number of interacting nucleons, can also improve the understanding of ZDC energy measured in $e$+A collisions at the future Electron-Ion Collider, where it was proposed as a possible centrality tag~\cite{Zheng:2014cha} to map out the impact parameter dependence on nPDFs.
A similar model, using a proton instead of a vector meson in the scattering with the nucleus, can be used for the nuclear breakup in \pxPb\ interactions. %

Given the importance of CF effects on centrality-dependent measurements of hard-scattering rates in \pxPb\ collisions, it is of great interest to study the correlation between the \kin\ of the hard-scattered partons, the neutron energy at very forward rapidity after nuclear breakup, and the forward calorimeter event activity in these events.
This inverts the typical approach by directly measuring the quantities used for centrality estimates as a function of jet kinematics, %
providing the first quantitative investigation of global observables in the presence of CF effects.

This Letter presents the first characterization of very forward energy in dijet events in \pxPb\ collisions at LHC energies, using 56 \NB\ of \pxPb\ data collected by ATLAS during 2016 at a nucleon--nucleon center-of-mass energy (\SNN) of 8.16~\TeV.
The energy deposited in the ZDC (\EZDC) and the transverse energy ($E_\mathrm{T}$) recorded in the FCal (\ETPB), both measured on the Pb-going side, %
are reported as a function of
\xp.
The normalized \EZDC\ and \ETPB\ distributions
and the means of these distributions, \AZDC\ and \EATPB, are analyzed to characterize the evolution of forward energy production in response to changes in the proton configuration.
The correlation between \ETPB\ and \EZDC\ is also investigated to provide insights about the underlying physics linking event activity and nuclear breakup dynamics in \pxPb\ collisions.

Following Ref.~\cite{ATLAS:2023zfx},
dijets are defined using the two highest transverse momentum\footnote{ATLAS uses a right-handed coordinate system with its origin at the nominal interaction point (IP)
in the center of the detector and the \(z\)-axis along the beam pipe.
The \(x\)-axis points from the IP to the center of the LHC ring,
and the \(y\)-axis points upwards.
Polar coordinates \((r,\phi)\) are used in the transverse plane,
\(\phi\) being the azimuthal angle around the \(z\)-axis.
The pseudorapidity is defined in terms of the polar angle \(\theta\) as \(\eta = -\ln \tan(\theta/2)\) and is proportional to the rapidity
$ y = \frac{1}{2} \ln \left( \frac{E + p_z}{E - p_z} \right) $ in the relativistic limit.  In the analysis the pseudorapidity sign is determined by the nuclear species, protons travel towards positive rapidities, while ions travel towards negative rapidities.
The rapidity in the proton--nucleon center-of-mass frame is shifted by +0.465 units compared with the ATLAS frame.
Angular distance is measured in units of $\Delta R\equiv \sqrt{ (\Delta \eta)^2 + (\Delta \phi)^2 }$. }
jets in a given collision, measured over a wide range of transverse momentum, \PT, and center-of-mass (CM) rapidity, $y^\mathrm{CM}$, to access a broad \xp\ phase space. The value of \xp\ is estimated from the final-state \kin\ of the two highest \PT\ jets in each event by using
\begin{equation}
\label{eq:xp}
x_p =
\frac{ p_{\TN{T},1}e^{y^\TN{CM}_{1}}
+
p_{\TN{T},2}e^{y^\TN{CM}_{2}}}{\sqrt{s_{_\text{NN}}}},
\end{equation}
where $p_{\TN{T},1}$ ($p_{\TN{T},2}$) and $y^\TN{CM}_{1}$ ($y^\TN{CM}_{2}$) denote the transverse momentum and the rapidity of the leading (sub-leading) jet.
At leading order in a perturbative approach, Equation~\ref{eq:xp} is the Bjorken-x of the scattered parton in the proton~\cite{PhysRevLett.123.102001}.  Direct estimates of the parton level kinematics through jet observables have been done in dijet measurements at the LHC~\cite{HION-2015-02} and RHIC~\cite{PhysRevLett.107.172301}.
In simulations, the \xp\ estimate based on final-state jet kinematics was found to be, on average, 6--8\% lower than the true parton-level value in the phase space covered by the analysis. The relative resolution of this estimate, $\sigma_{x_p^{\mathrm{reco}}/x_p^{\textsc{PYTHIA}}}$, is around 10\% for the lowest \xp\ events, decreasing to 4\% with increasing \xp.

In this analysis, \xp\ is unfolded for experimental effects due to the finite precision of jet reconstruction, and reported at the generator level.
\EZDC\ and \ETPB\ are not unfolded for experimental effects, as was done when used for centrality selections in previous analyses~\cite{ATLAS:2014cpa, ATLAS:2023zfx, ATLAS:2022iyq}.
These results present a novel approach to study geometry in proton--nucleus collisions and provide new insights into the role of partonic proton configurations and color fluctuations in the dynamics characterizing these reactions.
%


%

%
\section{ATLAS detector}
\label{sec:detector}
The measurement presented in this Letter is performed using the ATLAS inner detector, calorimeter, ZDC, trigger, and data acquisition
systems~\cite{PERF-2007-01}.

The inner detector measures charged particles within the pseudorapidity interval $|\eta| < 2.5$ using a combination of silicon pixel detectors,
silicon microstrip detectors (SCT), and a straw-tube transition radiation tracker (TRT), all immersed in a 2~T axial magnetic field.
Each of the three detectors is composed of a barrel and two symmetric endcap sections. The pixel detector barrel section is composed of four layers including
the insertable B-layer~\cite{ATLAS-TDR-19,PIX-2018-001}, and each endcap consists of three disks placed symmetrically on each side of the interaction region. The SCT barrel section contains four layers of modules with sensors on both sides,
and each endcap consists of nine layers of double-sided modules with radial strips. The TRT contains layers of staggered straws interleaved with
the transition radiation material.  Candidate collision vertices are reconstructed using inner detector tracks. The vertex with the highest track $\Sigma p_\mathrm{T}^2$ is defined as the primary vertex.

The calorimeter system consists of a sampling liquid-argon (LAr) electromagnetic (EM) calorimeter covering $|\eta| < 3.2$, a steel-scintillator sampling hadronic calorimeter covering $|\eta| < 1.7$, LAr hadronic calorimeters covering $1.5 < |\eta| < 3.2$,
and two LAr~FCals covering $3.2 < |\eta| < 4.9$.
The EM calorimeters are segmented longitudinally in shower depth into three layers with an additional presampler layer covering $|\eta| < 1.8$.
The hadronic calorimeters have three sampling layers longitudinal in shower depth for $|\eta| < 1.7$ and four sampling layers for $1.5 < |\eta| < 3.2$.
During the 2016 \pxPb\ data taking, a sector of the hadronic endcap calorimeter (HEC), corresponding to
$-3.2 < \eta < -1.5$ and
$-\pi < \phi < -\pi/2$ was disabled.
The FCal is composed of calorimeter towers segmented with fine granularity in $\Delta x\times \Delta y$, corresponding approximately to $\Delta \eta \times \Delta \phi = 0.15 \times 0.15 (0.3\times 0.3)$ at $\eta = 3.5 (4.5)$~\cite{PERF-2014-07}.
The EM energy scale is used to measure the energy in each tower ($E_\mathrm{tower}^\mathrm{EM}$).
The EM scale properly corrects the energy deposited by photons and electrons, but does not include any compensation to correct for the differences between EM and hadronic showers, as is standard for FCal measurements that are used as geometry estimators in \pxPb\ \cite{ATLAS:2014cpa, ATLAS:2023zfx, ATLAS:2022iyq} and \PbxPb\ \cite{HION-2022-14} measurements.
The ATLAS~ZDC consists of two detectors located in absorbers $\pm$140~m from the ATLAS interaction point.
Each detector is a sampling calorimeter made up of four modules each with a material budget of 1.14 hadronic interaction lengths.
The modules are made of layers of tungsten plates with quartz rods interspersed between them.
They measure forward-going neutral particles with $|\eta|>8.3$, primarily neutrons from nuclear breakup, which carry the original per-nucleon beam energy up to an additional smearing from nuclear Fermi momentum, and very forward neutral particles produced in the interaction.

To have sufficient experimental resolution to reconstruct a single neutron peak while simultaneously measuring calorimeter deposits from 80 or more neutrons from in-time pileup\footnote{In-time pileup refer to events containing multiple simultaneous \pxPb\ collisions.}, the ZDC requires a large dynamic range from its front-end electronics.  Experimentally, this is accomplished by configuring the ZDC with a dual gain readout system, where a high-gain analog-to-digital converter (ADC) samples with enough granularity to precisely measure relatively smaller energy deposits, while a low-gain module ensures that the ZDC does not saturate in events with sizable contamination from in-time pileup.  This dual-gain readout is combined at the reconstruction level to give a single measurement of \EZDC. In addition, having a reference for each signal allows for a data-driven correction of ADC non-linear behavior, applied at the level of the reconstruction.

ATLAS uses a two-level trigger system. The first-level trigger (Level 1) is hardware-based and implemented with custom electronics.
It is followed by the software-based high-level trigger (HLT)~\cite{TRIG-2016-01}.  An extensive software suite~\cite{SOFT-2022-02-Published} is used in the reconstruction and analysis of real and simulated data, in detector operations, and in the trigger and data acquisition systems of the experiment.


%

%
\section{Data selection and Monte Carlo simulations}
For the 2016 \pxPb\ run at \SNN~=~8.16~\TeV, the LHC beam energy configuration was asymmetric between the protons (6.5~\TeV), and the Pb nuclei (2.56~\TeV/nucleon), resulting in a rapidity shift of the center of mass by 0.465 units toward the $p$-going direction.
The data were collected over two running periods characterized by interchanging species between beam directions.
In the first period of the data-taking, when Pb~ions %
circulated clockwise %
in the LHC, a total integrated luminosity of 56~\NB\ was collected.
The data analyzed in this Letter are comprised solely of the first data-taking period, referred to as the \pxPb\ orientation.
The second data-taking period is ommited since the presence of a detector from the LHCf experiment~\cite{Adriani:2135330} precludes a precise measurement of the spectator neutron energy.
The positive \ETA\ direction used in the analysis is defined by the proton-going direction.

The data used
satisfy detector and data-quality requirements, and contain at least one reconstructed vertex and two reconstructed jets.
A set of fully efficient central and forward single-jet triggers~\cite{TRIG-2016-01}, characterized by different \PT\ thresholds, was chosen to provide full \PT\ coverage over the phase space of this analysis. %

Jets are formed from calorimeter towers and are reconstructed using the \antikt algorithm~\cite{Cacciari:2008gp,Fastjet} with E-scheme recombination and \RFour, as was done in previous ATLAS jet measurements in \pxPb~\cite{ATLAS:2022iyq, ATLAS:2023zfx}.
This analysis considers events where the two highest \PT\ jets have reconstructed $p_{\TN{T,1}}>40$~\GeV, $p_{\TN{T,2}}>30$~\GeV\ and $-2.8 < \eta_{1},\eta_{2} < 4.5$, respectively.
Sub-leading jets have a lower \PT\ threshold to allow for some momentum imbalance in the reconstructed dijet at low \PTA.
The asymmetric \ETA\ requirement is imposed to avoid jets biasing the centrality determination in the Pb-going FCal.
Events with either the leading or the sub-leading jet reconstructed in the acceptance of the disabled HEC region are discarded in both data and Monte Carlo (MC) simulations.
To define a rejection criterion for the analysis, the disabled region is increased by an additional 0.4 margin in both pseudorapidity and azimuthal angle, as done in previous jet analyses using these datasets~\cite{ATLAS:2023zfx,ATLAS:2022iyq,HION-2017-06}.
In this way, jets with constituents affected by the disabled HEC are not considered and an efficiency correction is applied (see Section~\ref{sec:analysis}).

Centrality is defined using
\ETPB~\cite{ATLAS:2015hkr,ATLAS:2019vcm,ATLAS:2023zfx}.
To reject contamination from UPC events, the centrality range 0--90\% is considered.
The contribution of UPC jet production in these events was studied by applying an additional selection based on the rapidity gap~\cite{HION-2018-27} and is found to be negligible.
In-time pileup events are rejected by requiring that no non-primary vertices have more than six associated tracks\footnote{The average interactions per bunch crossing, $\bar\mu$, varied throughout the \pxPb\ data-taking. A majority of the sampled luminosity had a modest pileup profile $ 0.1 < \bar \mu < 0.3$. Low pileup samples $\bar \mu < 0.01$ were taken to study the impact of pileup on the data.}.
This removes approximately 12\% of events from the sample.
All these selections are consistent with the \pxPb\ analysis presented in Ref.~\cite{ATLAS:2023zfx}

The ZDC ADCs sample their own baseline before every event.
This baseline is sensitive to the long tails of preceding \pxPb\ interactions with significant ZDC activity.
By rejecting events with baseline values elevated relative to an acceptable range taken from a dedicated zero pileup sample, events with an out-of-time\footnote{Out-of-time pileup refers to events preceded by another \pxPb\ collision where a detector measures signals originating from the preceding event in the sampling window of the current event.} pileup contribution that can affect measurements in the FCal are excluded.
This requirement removes approximately 8\% of events, after the nominal pileup cut and the centrality selection.
This requirement is agnostic to the kinematic properties of the considered event, as it only depends on the process in a previous event.

MC simulations are used to evaluate the performance of the detector and analysis procedure, and to correct the measured distributions for detector effects. The detector response in all MC samples is simulated using \textsc{Geant4}~\cite{Agostinelli:2002hh,SOFT-2022-02-Published}.
The disabled HEC sector is replicated in the simulation.
The \pxPb\ MC sample makes use of dijet events from 8.16~\TeV\ $pp$ collisions, including the boost in rapidity relative to the lab frame that is in the data, simulated by \pythiaeight~\cite{Sjostrand:2014zea} with the A14 set of tuned parameters~\cite{ATL-PHYS-PUB-2014-021} and the~\textsc{NNPDF2.3lo} parton distribution functions~\cite{Ball:2012cx}.
Events from the \pythiaeight\ dijet sample are overlaid with events from a dedicated \pxPb\ data sample collected using MB triggers.
The overlay procedure combines the \pythiaeight\ and data events during the digitization step of simulation.
The \ETPB\ from the \pythiaeight\ event is mostly insensitive to the \xp\ estimated from the dijet kinematics~\cite{HION-2015-02}.
An event-by-event reweighting procedure is applied to the MC overlay sample, such that the resulting \ETPB\ distribution better matches that of the dijet data sample.

%
%

%

%
%
%

%
%
%


%

%
\section{Analysis}
\label{sec:analysis}
The value of \ETPB\ is defined by a sum over the projection of each $E_\mathrm{tower}^\mathrm{EM}$ into the transverse plane, for the towers in the pseudorapidity hemisphere facing the Pb ion beam.
This definition of \ETPB\ is consistent with the characterization of centrality used in previous ATLAS \pxPb\ analyses, for example in Refs.~\cite{HION-2013-08, ATLAS:2023zfx}.

\begin{figure}[t]
\centering
\includegraphics[width=0.5\linewidth]{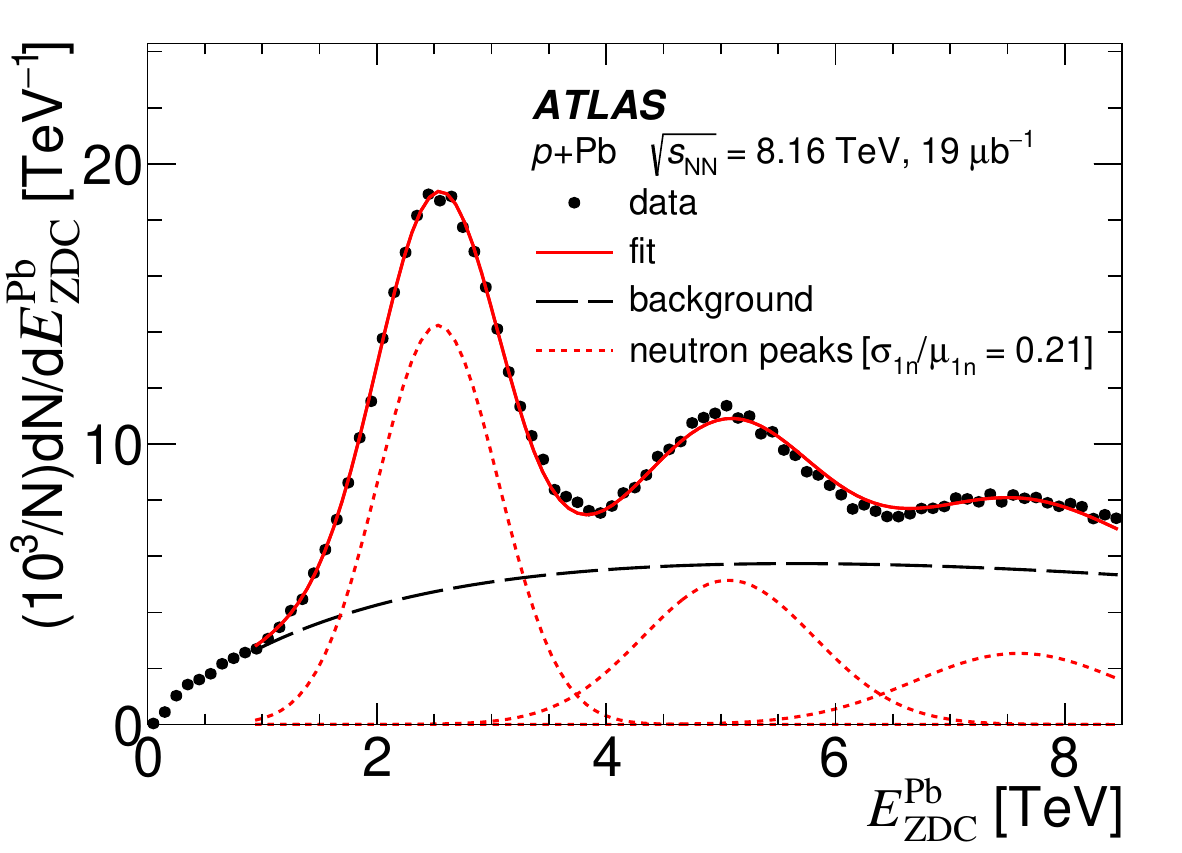}
\caption{Low energy region of the calibrated ZDC spectrum for events triggered using the proton-going ZDC\@. The fit result is shown by the solid line. All the fit components are displayed using dashed lines.
} %
\label{fig:single_neutron_peak}
\end{figure}

In a typical \pxPb\ event involving a hard scattering, most of the energy deposited into the Pb-going ZDC originates from spectator neutrons evaporating off the struck nucleus.
A peaked structure appears in the low-energy region of the ZDC spectrum due to the detector's response to a small number of neutrons, which can be clearly distinguished.
An example of the ZDC energy spectrum and peak fitting result is displayed in Figure~\ref{fig:single_neutron_peak}.
The first peak in this structure represents the ZDC response to the energy of a single spectator neutron.
The absolute energy response of the ZDC modules is calibrated by a Lagrangian optimization procedure that sets the mean of the single neutron peak equal to the nominal per nucleon beam energy and adjusts calibration factors in each layer to minimize the width of the single neutron peak.
The calibration procedure is repeated throughout the run to account for intra-run changes in the detector response.
Following calibration, a simultaneous fit is performed to the first three neutron peaks, as well as a background term.
Background sources include neutral pions, photons, and beam--beam backgrounds.
The functional form of the background term is chosen to facilitate fit convergence in the 1--3 neutron range.
On physical grounds this contribution should vanish at zero energy and at large energies.
These fits are used to validate the energy scale set by the calibration.
The fit procedure is applied on a per-run basis, to account for changes in the detector response during data-taking.

The width of the single neutron peak originates from two separate sources.
First, the ZDC is a sampling calorimeter that has a finite experimental resolution. %
Second, the contribution from nuclear Fermi momentum smears the neutron energy away from the nominal per-nucleon beam energy.
The use of a Gaussian distribution to model the calorimeter response to a single neutron is motivated by the assumption that the modification from the convolution of these two sources is roughly symmetric.
Below 1~\TeV, the detector is primarily observing a mix of low-energy photons and beam backgrounds, with a negligible fraction from single neutrons. The ZDC amplitude reconstruction procedure also requires pulse heights well above detector noise that limit its precision below 1~\TeV. Therefore, below that threshold, the measured energy is treated as zero, i.e., compatible with zero neutrons showering in the ZDC.

A residual pileup subtraction is applied to the \EZDC\ distribution, to remove in-time pileup contributions from interactions that are not rejected by the vertex requirement, and that are characterized by no significant pileup energy deposit in the FCal. This procedure removes approximately 0.19\% of the events from the sample.

The jet reconstruction performance for this data sample was scrutinized in detail in a previous analysis, see Ref.~\cite{ATLAS:2023zfx} for all the relevant details.
To correct for detector effects and bin migration, dijet events are unfolded in \xp\ using a two-dimensional Bayesian procedure~\cite{DAgostini:1994fjx}, implemented within the RooUnfold package~\cite{Adye:2011gm} and validated by closure tests.
The value of \xp\ at particle level is estimated with Eq.\ref{eq:xp} using the leading and sub-leading particle-level jets.
\ETPB\ and \EZDC\ are individually propagated through the unfolding procedure with fully diagonal migration matrices, preserving the event-level correlation between the hard-scattering \kin\ and each of the calorimetric energies.
An efficiency correction is included in the unfolding to account for reconstructed jets that migrate out of the measurement phase space at the detector-level due to energy resolution effects, as well as to the disabled HEC region.

Each response matrix is reweighted at the event level by the ratio of reconstructed data to reconstructed simulation, as a function of \xp, such that the simulated spectrum better matches the shape of the data.
The statistical uncertainties in the resulting \ETPB\ and the \EZDC\ distributions, binned as a function of \xp, are evaluated using a bootstrapping method~\cite{ATL-PHYS-PUB-2021-011}.


%

%
\section{Systematic uncertainties}
This measurement unfolds experimental effects in the dijet measurements to access \xp\ at the particle level.
This approach is subject to systematic uncertainties associated with the jet energy scale (JES) and jet energy resolution (JER), the unfolding procedure, and an uncertainty due to a sector of the HEC being disabled for the running period.
Measurements of \EZDC\ and \AZDC\ have additional uncertainties applied to the energy scale of the ZDC\@.
Since the ZDC energies are not unfolded, these uncertainties are evaluated at the reconstructed level. Other sources of systematic uncertainty, such as those related to the removal of the residual pileup in the ZDC, were found to be negligible. %
For each source of systematic uncertainty, except for the uncertainties in the disabled HEC sector and the ZDC energy scale, the entire analysis is repeated by changing the response matrix according to the systematic variation.
The difference between the nominal measurement and that obtained with the systematic variation is taken as the systematic uncertainty.

The treatment of the uncertainties related to the determination of \xp\ via jets, namely the uncertainties in the JES, JER, and HEC exclusion, is identical to the procedure outlined in Ref.~\cite{ATLAS:2023zfx} and the discussion of these uncertainties is minimized in this Letter.
The JES and JER correspond to the mean and variance of the $p_{\TN{T}}^\TN{reco}/p_{\TN{T}}^\TN{gen}$ distributions, where $p_{\TN{T}}^\TN{gen}$ denotes the transverse momentum of the matched generator-level jet in simulation.
The HEC exclusion uncertainty evaluates the sensitivity to increasing the exclusion region by 0.1 in both azimuth and pseudorapidity, and repeating the analysis procedure.

The systematic uncertainty in the unfolding procedure is related to the sensitivity to the choice of the prior distribution, i.e., the response matrix reweighting described in Section~\ref{sec:analysis}.
To determine the sensitivity of the results to the reweighting procedure, a new set of response matrices was generated without reweighting at the event level, and the full analysis procedure was repeated.
An uncertainty associated with the ZDC energy stability is assigned to quantify the run-by-run variations in the single neutron peak position extracted from fits like the one shown in Figure~\ref{fig:single_neutron_peak}.
The residual between the nominal per-nucleon beam energy (2.56~\TeV) and the extracted peak position, after allowing for 1$\sigma$ deviations from the stability, is taken as an uncertainty in the ZDC energy scale.
This approach was chosen to avoid overestimating the uncertainties assigned to the ZDC measurements. This residual is intrinsically linked to the finite time-resolution of the calibration procedure described in Section~\ref{sec:analysis}.
A systematic uncertainty is also assigned to the ZDC non-linear correction mentioned in Section~\ref{sec:detector}. This uncertainty is determined by calculating the difference between \EZDC\ with and without the non-linear correction applied, across the whole spectrum of \EZDC\ values.  Akin to the previous ZDC uncertainties, an additional systematic uncertainty is assigned to the stability of this correction and any residual scale variation.

The stability of the FCal in measuring \ETPB\ across the run was scrutinized and the associated uncertainty is found to be negligible compared with other systematic uncertainties assigned.
The variations for each systematic uncertainty were then added in quadrature to produce the total systematic uncertainties in each data point.

The \EZDC\ total systematic uncertainties are dominated by the ZDC energy stability term, which is within 7\% for most bins considered.  The uncertainty grows towards high \EZDC\ values and at the edges of the \xp\ measurement range, where it reaches up to 30\%.

The total systematic uncertainties in the \ETPB\ distributions in various \xp\ bins is dominated by contributions from the JES, JER, and unfolding prior components.  It is within 10\% for most bins, increasing at high and low \ETPB\ values and at the edges of the \xp\ measurement range, where it gets up to 20\%.
The total systematic uncertainties in the distributions of \AZDC\ and \EATPB\ are below 2\% in all the \xp\ bins except for the lowest, where it reaches 3\%.
When ratios are constructed, the JES, JER, HEC, and ZDC uncertainties are treated as fully correlated between measurement interval, while the prior uncertainty is treated as uncorrelated.
%

%
%
%
%
%
%

%
%
%
%
%
%


%

%
%
%
%
\section{Results}

\begin{figure}[h!]
\centering
\begin{subfigure}[c]{0.48\linewidth}
\includegraphics[width=0.97\linewidth,trim={2mm 2mm 2mm 10mm}]{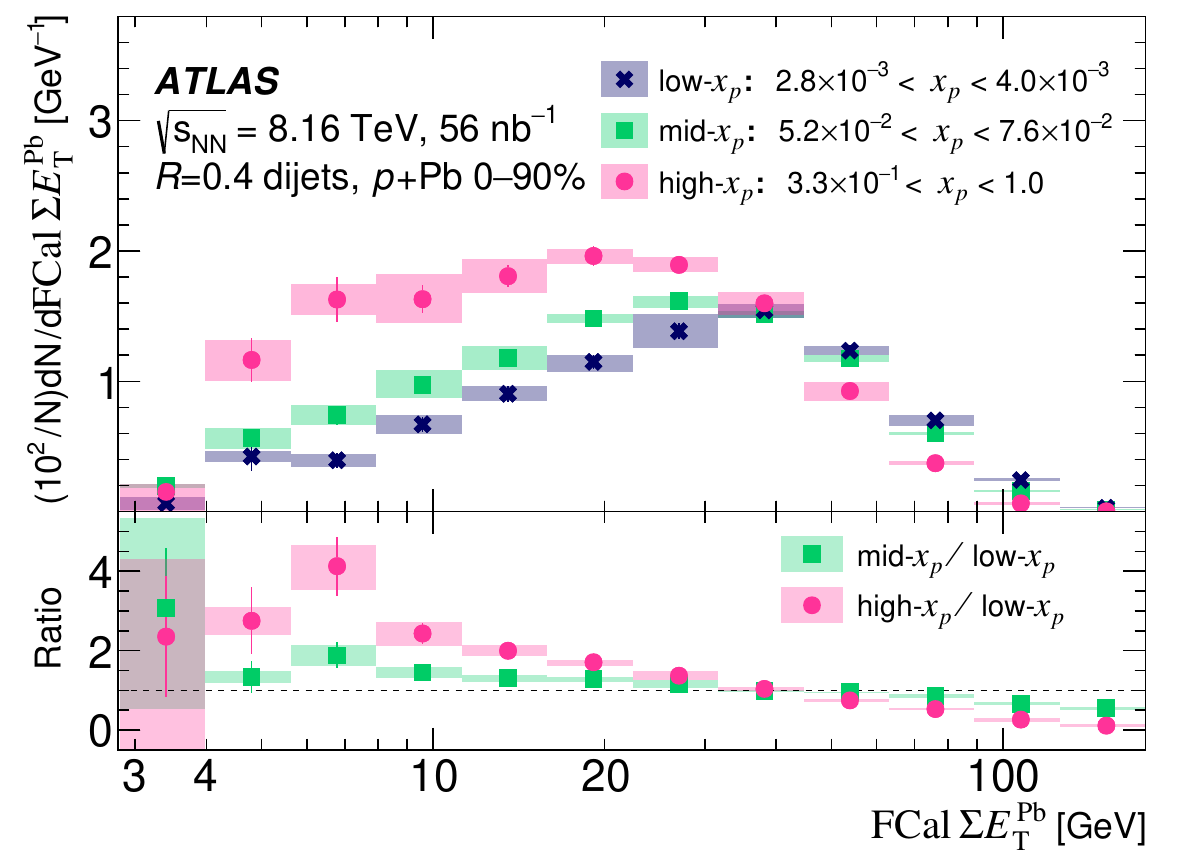}
\caption{}
\label{fig:unf_select_xp_fcal}
\end{subfigure}
\begin{subfigure}[c]{0.48\linewidth}
\includegraphics[width=0.97\linewidth,trim={2mm 2mm 2mm 10mm}]{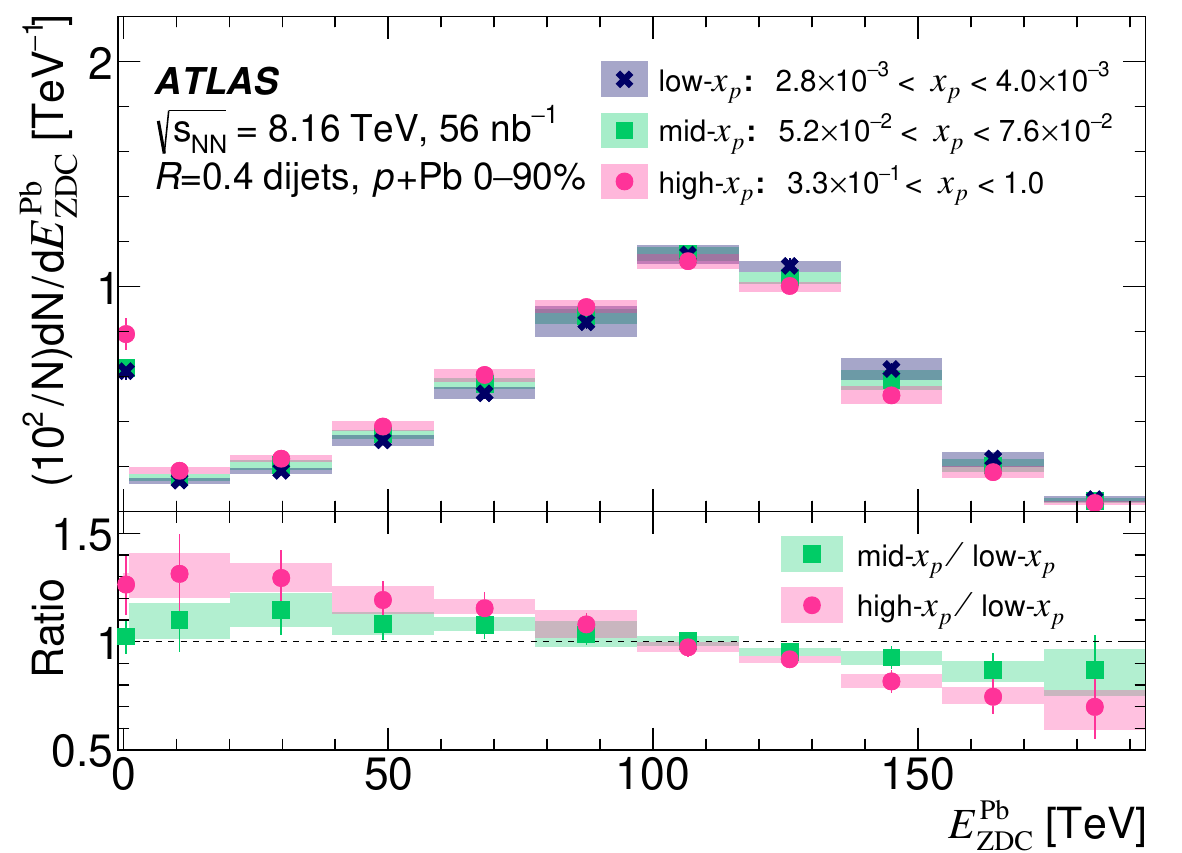}
\caption{}
\label{fig:unf_select_xp_zdc}
\end{subfigure}
\caption{(a) \ETPB\ and (b) \EZDC\ distributions plotted for low-, middle-, and high-\xp\ selections. These distributions are normalized to display the change of shape with \xp. The bottom panel displays the ratio of the middle- and high-\xp\ selection over the low-\xp\ selection. In sub-figure (b) the energy below the ZDC detection threshold (e.g., formally assigned as zero by ZDC reconstruction) is reported in a dedicated bin.
Please note that results in each bin are divided by the bin width. This is relevant in sub-figure (b) comparing the zero energy bin to the rest of the distribution.
Vertical error bars represent the statistical uncertainty, and the shaded areas represent the systematic uncertainty.  The width of the shaded area indicates the bin width.} %

\end{figure}

Normalized distributions of \EZDC\ and \ETPB\ in dijet events are presented in intervals of \xp
, estimated by using Eq.~\ref{eq:xp}, chosen to allow a direct comparison with the results of Ref.~\cite{ATLAS:2023zfx}. The highest and lowest bins are wider to accrue more statistics.
Figure~\ref{fig:unf_select_xp_fcal} shows a comparison of \ETPB\ distributions in well-separated intervals of \xp.
The lower panel of Figure~\ref{fig:unf_select_xp_fcal} shows the ratio of the middle- and high-\xp\ \ETPB\ distributions to that in low-\xp\ dijet events.
These ratios show that, relative to a lower-\xp\ selection, higher-\xp\ events have a \ETPB\ distribution that is significantly different and shifted towards lower energies. This observation represents the first direct confirmation of the event activity bias first observed by ATLAS in jet measurements~\cite{ATLAS:2014cpa,ATLAS:2023zfx}.

The distributions of \EZDC\ are presented, in the same fashion as Figure~\ref{fig:unf_select_xp_fcal}, in Figure~\ref{fig:unf_select_xp_zdc}.
The lower panel of Figure~\ref{fig:unf_select_xp_zdc} shows the ratio of the middle- and high-\xp\ \EZDC\ distributions to that in low-\xp\ dijet events.
The ratio reveals a significant difference between the \EZDC\ distribution in high-\xp\ and low-\xp\ events, where the high-\xp\ distribution has a higher fraction of events with fewer forward neutrons.
This conclusion for the ZDC is qualitatively similar to, but substantially weaker than, the event activity bias in the FCal observed in Figure~\ref{fig:unf_select_xp_fcal}.

The first bin in all the %
\EZDC\ distributions of Figure \ref{fig:unf_select_xp_zdc} is set to 0--1~\TeV, to isolate events that are compatible with zero neutrons in the detector.
The fraction of events in these bins is smaller than 1\%.  These distributions are normalized to the bin width, introducing the visual jump between the first bin and the rest of the distribution.  Furthermore, in Ref.~\cite{PhysRevC.107.064902} ALICE observes a small fraction of hadronic \PbxPb\ collisions with neutron topologies consistent with zero neutrons in one ZDC side.  Extending this to the \pxPb\ case, one might expect a small fraction of hadronic events to also be characterized by zero breakup neutrons.

These observations of shifting distributions, for both \EZDC~and~\ETPB, are consistent with a reduction in the number of interactions between the proton projectile and the nucleons in the nucleus with increasing \xp. The relative magnitude of the shift in the distribution of calorimeter energies is significantly larger in the \ETPB\ case compared with the \EZDC\ case.
\pagebreak

\FloatBarrier
\begin{figure}[ht]
\centering
\begin{subfigure}[c]{0.48\linewidth}
\includegraphics[width=0.97\linewidth,trim={1mm 1mm 5mm 3mm},clip]{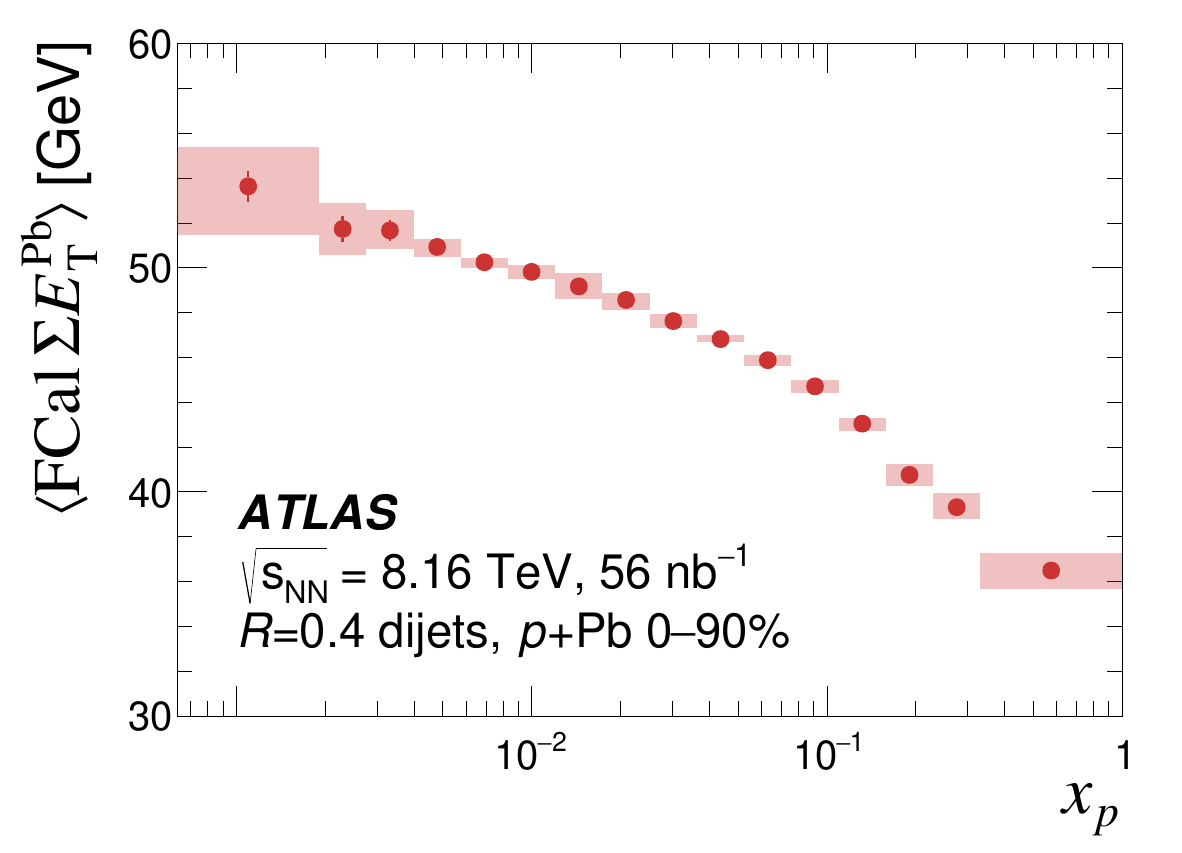}
\caption{}
\end{subfigure}
\begin{subfigure}[c]{0.48\linewidth}
\includegraphics[width=0.97\linewidth,trim={1mm 1mm 5mm 3mm},clip]{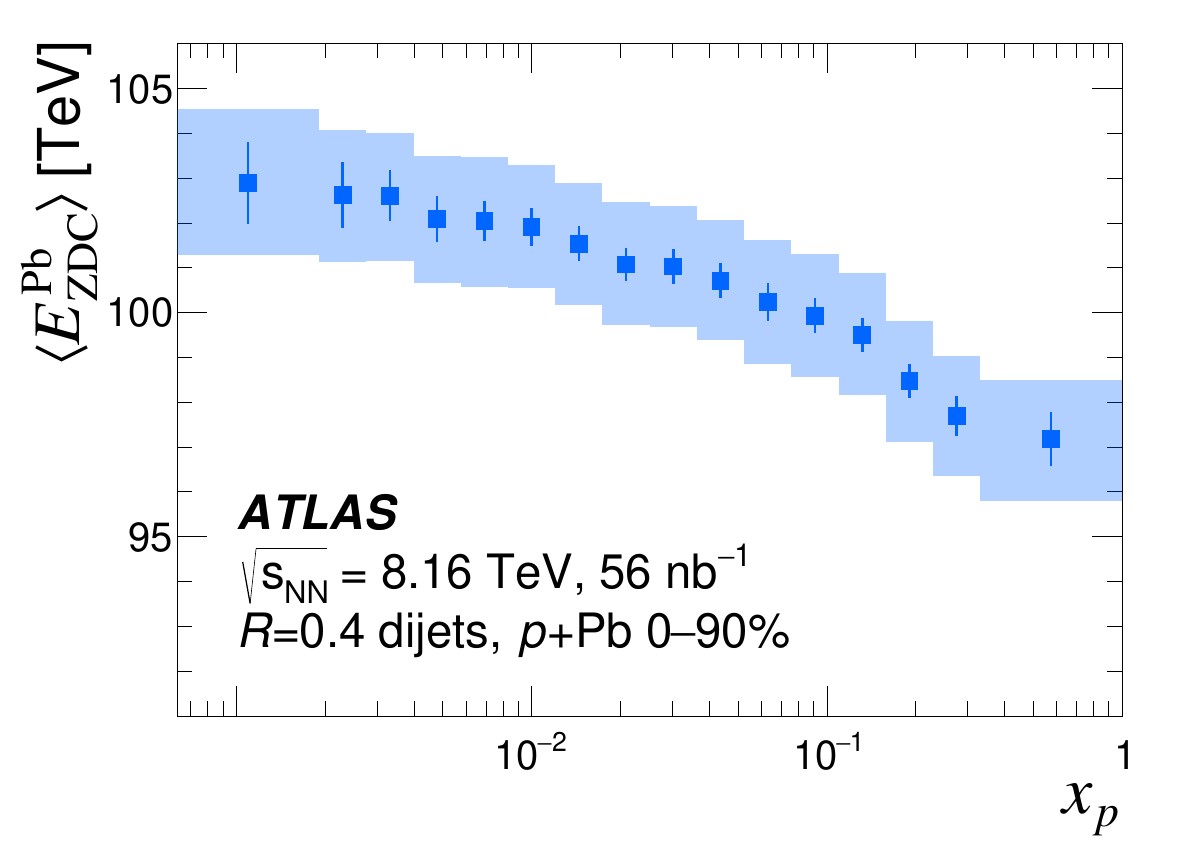}
\caption{}
\end{subfigure}
\caption{(a) \EATPB\ and (b) \AZDC\ both as a function of \xp.
Vertical error bars represent the statistical uncertainty, and the shaded areas represent the systematic uncertainty.  The width of the shaded area indicates the bin width.
}
\label{fig:unfresult_avg}
\end{figure}

\begin{figure}[ht]
\centering
\includegraphics[width=0.9\linewidth,trim={1mm 1mm 2mm 4mm},clip]{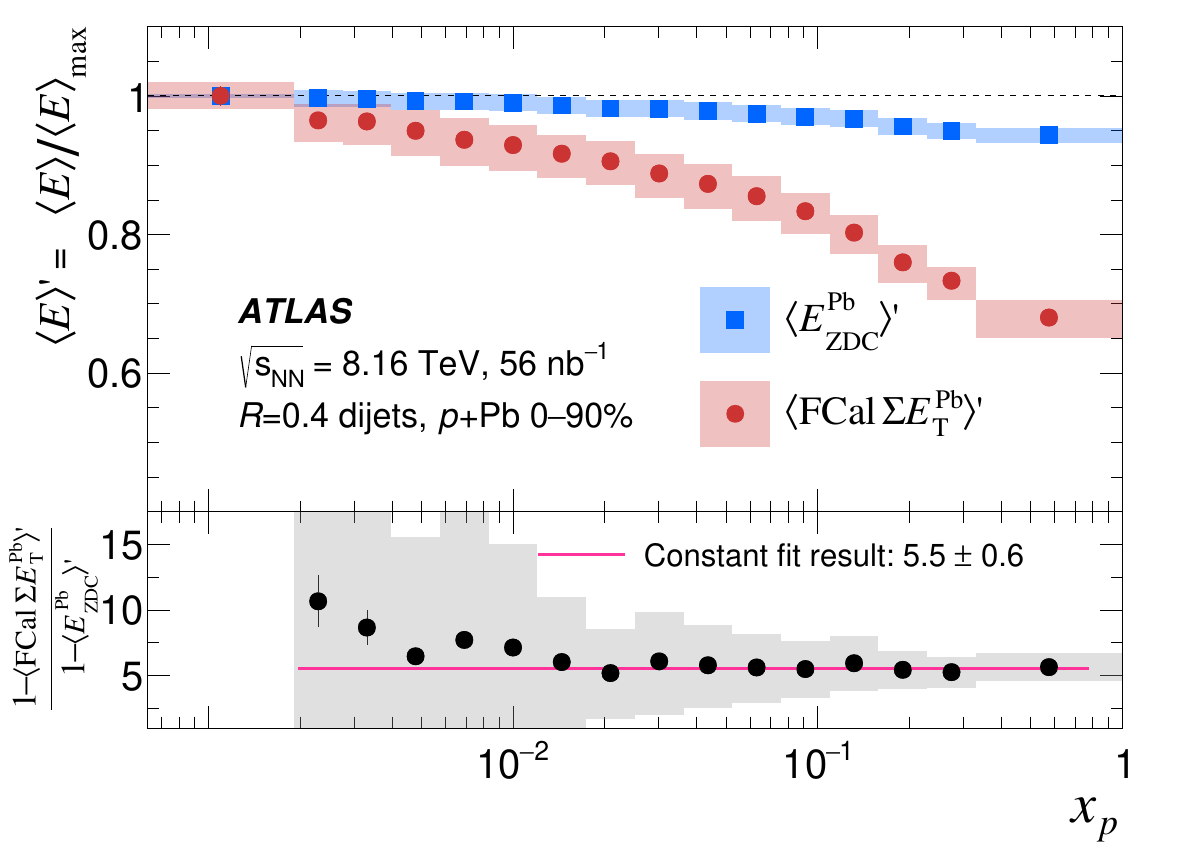}
\caption{\AZDC\ as a function of \xp\ (square) compared with \EATPB\ as a function of \xp\ (circle). Both plots are normalized to their maximum value, to display both quantities on the same scale.
The bottom panel shows the ratio of the relative change in \EATPB\ over the relative change in \AZDC, fit to a constant function (solid line).
Shaded areas represent the systematic uncertainty.  The width of the shaded area indicates the bin width. The statistical error on the $\langle E \rangle_{\mathrm{max}}$ is included in the systematic band as it is correlated bin-to-bin in \xp\, for all bins except for the lowest \xp\ bin.
The vertical error bars represent the uncorrelated statistical uncertainty. }
\label{fig:unfresult_rel_cal_change}
\end{figure}

To better understand the correlations between \xp\ and the energy (transverse energy) detected in the ZDC (FCal), the mean values of the distributions are presented as a function of \xp\ in Figure~\ref{fig:unfresult_avg}. Both curves exhibit a decrease as a function of \xp, more pronounced in the region \xp~$\gtrsim 0.02$, corresponding to the onset of effects attributed to small proton size configurations in Ref.~\cite{ATLAS:2023zfx}. The energy accumulated in the Pb-going side of the ZDC decreases by up to $\sim$5\% in this region, while a decrease of up to 40\% is observed for the FCal.
The observed decrease of $\sim$5\% in the \AZDC\ corresponds to, on average, an energy detected in the ZDC corresponding to two fewer beam-energy neutrons.
Such a trend is qualitatively compatible with a reduced number of interactions for small proton configurations resulting in a smaller number of breakup neutrons, as suggested in Ref.~\cite{Alvioli:2024cmd}.

Figure~\ref{fig:unfresult_rel_cal_change} shows the comparison between the rates of the \xp-dependent decrease between \EZDC\ and \ETPB, both normalized to their maximum values. The lower panel shows the ratio of the relative change in \EATPB\ over the relative change in \AZDC\ as a function of \xp.
The ratio remains approximately constant, as shown by the fit in the Figure.
For \xp\ bins~$\lesssim 0.02$, the systematic uncertainty in the ratio becomes large despite cancellations from correlated sources of uncertainty, as even small deviations in \AZDC\ in this region where \AZDC\ is close to to its maximal value can have a large effect on the ratio denominator.  However, the ratio remains consistent with the constant fit within the large systematic uncertainties.
This result shows that, in \pxPb\ collisions, the energy deposited in the ZDC is about six times less sensitive to the hard process kinematics than the transverse energy at FCal rapidities.
\FloatBarrier
The correlation between the transverse energy produced by the interactions of the participants and the energy deposited by the spectator neutrons in the ZDC in \xp\ selections, measured at the reconstructed level, is also studied. The Pearson correlation coefficient between \EZDC\ and \ETPB, calculated for events in each \xp\ bin, decreases by 3\% from the high-\xp\ to low-\xp.
This finding suggests that, in a smaller proton configuration, the forward transverse energy produced by the interactions of the participants is slightly less correlated with the number of spectator neutrons at very forward rapidities.

\begin{figure}[t]
\centering
\includegraphics[width=0.8\linewidth,trim={0mm 0cm 6mm 0mm},clip]{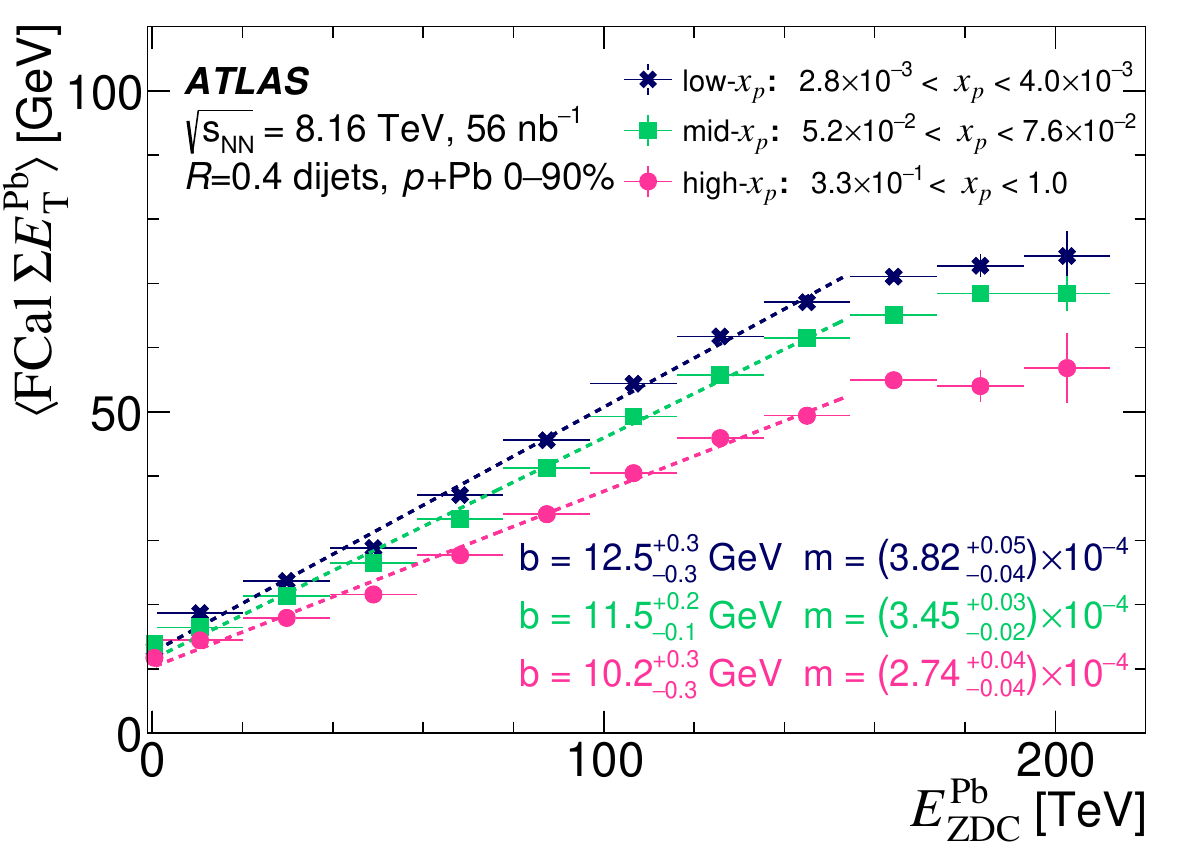}
\caption{\EATPB\ plotted as a function of \EZDC\ for three selections of \xp: a low, middle, and high selection. The data in each bin are fit to a linear form ($f(x) = m\cdot x + b$) to approximately describe the correlation between the two energies.  Fit parameters (and fit uncertainties) are reported on the figure.  The values of \xp\ are measured at the reconstructed level. %
}
\label{fig:fit_select_zdc_fcal}
\end{figure}
\FloatBarrier

To further study the evolution of the correlation between \ETPB\ and \EZDC\ with \xp, Figure~\ref{fig:fit_select_zdc_fcal} reports \EATPB\ as a function of \EZDC\ for three different \xp\ selections.
A
relationship of \EATPB\ as a function of \EZDC\
consistent with a linear dependence is observed in the range of 0--150~\TeV\
in \EZDC\ for all the \xp\ bins.
In Figure~\ref{fig:fit_select_zdc_fcal}, each distribution is fit to a linear function %
over that domain to approximately capture the underlying relation between the energies measured in the two calorimeters.
The data above 150~\TeV, corresponding to $\sim$60 neutrons, are not considered in the fit since visible deviations from the linear trend
start to appear.
This behavior is likely due to fluctuations in the nuclear breakup occurring in events characterized by the highest (most central) event activity (collision geometry).
The dependence of the linear fit's slope and intercept on the fit range was studied by systematically varying the upper limit of the fit range to cover $\pm$1 bin in \EZDC\ relative to the nominal choice of 150~\TeV.
The slope of the linear correlation between \EATPB\ and \EZDC\ decreases progressively moving from low to high-\xp\ values, an observation consistent with the \xp\ dependence of the Pearson correlation coefficient. Both these quantities are useful metrics for describing how the relationship between \EZDC\ and \ETPB\ changes with \xp.
However, a change in the number of neutrons does not directly imply a change in the transverse energy produced by the interactions of the participants at forward rapidity; they change as they both are correlated with the \pxPb\ collision geometry.

\clearpage


%

%
\section{Conclusion}

This Letter presents the characterization of the dependence of two event geometry estimators, used in the analysis of hard-scatterings in \pxPb\ collisions, on the initial-state kinematics. The transverse energy deposited in the ATLAS Forward Calorimeter and the forward energy recorded by the Zero-Degree Calorimeter in \pxPb\ collisions at \SNN~=8.16~\TeV\ are analyzed as a function of the estimated Bjorken-$x$ of the proton, \xp, accessed through dijet events detected using the ATLAS calorimeter.
Both \EZDC\ and \ETPB\ are shown to be sensitive to \xp, especially in the region \xp~$\gtrsim 0.02$, where previous ATLAS results from the same data~\cite{ATLAS:2023zfx} have shown strong evidence of event activity biases that can be interpreted in terms of color fluctuation effects related to the proton configuration at the moment of the hard-scattering~\cite{Alvioli:2017wou}.
The average \EZDC\ shows variations corresponding to two beam-energy neutrons between the high- and low-\xp\ selections considered in the analysis.
Still, compared with the transverse energy in the Pb-going FCal, the energy deposited in the Pb-going ZDC is found to be about six times more robust against variations in the event kinematics for measurements of hard process rate modifications.

Further studies on the correlation between the two geometry estimators as a function of \xp\ are also reported. This analysis exposes a linear, \xp-dependent, relation between \ETPB\ and \EZDC.
These results can be used to inform the simultaneous modeling of event activity and nuclear breakup in \pxPb\ collisions characterized by the presence of a hard-scattering.

The results presented in this Letter qualitatively support models that connect color fluctuations with neutrons evaporated in the nuclear breakup~\cite{Alvioli:2024cmd, Alvioli:2025ggv}. Access to the underlying physics linking hard-scatterings and nuclear breakup will also aid the understanding of nuclear evaporation in processes characterized by resolved photon exchange, such as dijet production in UPCs at the LHC and $e$+A collisions at the future Electron-Ion Collider.
%

%


%

%
%
%
%
%
\section*{Acknowledgements}
%

%

%
%

%
%

We thank CERN for the very successful operation of the LHC and its injectors, as well as the support staff at
CERN and at our institutions worldwide without whom ATLAS could not be operated efficiently.

The crucial computing support from all WLCG partners is acknowledged gratefully, in particular from CERN, the ATLAS Tier-1 facilities at TRIUMF/SFU (Canada), NDGF (Denmark, Norway, Sweden), CC-IN2P3 (France), KIT/GridKA (Germany), INFN-CNAF (Italy), NL-T1 (Netherlands), PIC (Spain), RAL (UK) and BNL (USA), the Tier-2 facilities worldwide and large non-WLCG resource providers. Major contributors of computing resources are listed in Ref.~\cite{ATL-SOFT-PUB-2026-001}.

We gratefully acknowledge the support of ANPCyT, Argentina; YerPhI, Armenia; ARC, Australia; BMWFW and FWF, Austria; ANAS, Azerbaijan; CNPq and FAPESP, Brazil; NSERC, NRC and CFI, Canada; CERN; ANID, Chile; CAS, MOST and NSFC, China; Minciencias, Colombia; MEYS CR, Czech Republic; DNRF and DNSRC, Denmark; IN2P3-CNRS and CEA-DRF/IRFU, France; SRNSFG, Georgia; BMFTR, HGF and MPG, Germany; GSRI, Greece; RGC and Hong Kong SAR, China; ICHEP and Academy of Sciences and Humanities, Israel; INFN, Italy; MEXT and JSPS, Japan; CNRST, Morocco; NWO, Netherlands; RCN, Norway; MNiSW, Poland; FCT, Portugal; MNE/IFA, Romania; MSTDI, Serbia; MSSR, Slovakia; ARIS and MVZI, Slovenia; DSI/NRF, South Africa; MICIU/AEI, Spain; SRC and Wallenberg Foundation, Sweden; SERI, SNSF and Cantons of Bern and Geneva, Switzerland; NSTC, Taipei; TENMAK, T\"urkiye; STFC/UKRI, United Kingdom; DOE and NSF, United States of America.

Individual groups and members have received support from BCKDF, CANARIE, CRC and DRAC, Canada; CERN-CZ, FORTE and PRIMUS, Czech Republic; COST, ERC, ERDF, Horizon 2020, ICSC-NextGenerationEU and Marie Sk{\l}odowska-Curie Actions, European Union; Investissements d'Avenir Labex, Investissements d'Avenir Idex and ANR, France; DFG and AvH Foundation, Germany; Herakleitos, Thales and Aristeia programmes co-financed by EU-ESF and the Greek NSRF, Greece; BSF-NSF and MINERVA, Israel; NCN and NAWA, Poland; La Caixa Banking Foundation, CERCA and AGAUR programs from Generalitat de Catalunya and PROMETEO and GenT Programmes Generalitat Valenciana, Spain; G\"{o}ran Gustafssons Stiftelse, Sweden; The Royal Society and Leverhulme Trust, United Kingdom; Eric and Wendy Schmidt Fund for Strategic Innovation, United States of America.

In addition, individual members wish to acknowledge support from Armenia: Yerevan Physics Institute (FAPERJ); CERN: European Organization for Nuclear Research (CERN DOCT); Chile: Agencia Nacional de Investigaci\'on y Desarrollo (ANID FONDECYT reg. 1230987, FONDECYT 1230812, FONDECYT 1240864, Fondecyt 3240661, Fondecyt Regular 1240721); China: Chinese Ministry of Science and Technology (MOST-2023YFA1605700, MOST-2023YFA1609300), National Natural Science Foundation of China (NSFC - 12175119, NSFC 12275265); Czech Republic: Czech Science Foundation (GACR - 24-11373S), Ministry of Education Youth and Sports (ERC-CZ-LL2327, FORTE CZ.02.01.01/00/22\_008/0004632), PRIMUS Research Programme (PRIMUS/21/SCI/017); EU: H2020 European Research Council (ERC - 101002463); European Union: European Research Council (BARD No. 101116429, ERC - 948254, ERC 101089007), European Regional Development Fund (HE COFUND GA No.101081355, ERDF), Horizon 2020 Framework Programme (MUCCA - CHIST-ERA-19-XAI-00), European Union, Future Artificial Intelligence Research (FAIR-NextGenerationEU PE00000013), Horizon 2020 (EuroHPC - EHPC-DEV-2024D11-051), Italian Center for High Performance Computing, Big Data and Quantum Computing (ICSC, NextGenerationEU), Marie Sklodowska-Curie Actions (ERC Horizon-2020, GA 956086, GAP-101168829); France: Agence Nationale de la Recherche (ANR-21-CE31-0013, ANR-21-CE31-0022, ANR-22-EDIR-0002, ANR-24-CE31-0504-01); Germany: Baden-Württemberg Stiftung (BW Stiftung-Postdoc Eliteprogramme), Deutsche Forschungsgemeinschaft (DFG - 469666862, DFG - CR 312/5-2); China: Research Grants Council (GRF); Italy: Istituto Nazionale di Fisica Nucleare (ICSC, NextGenerationEU), Ministero dell'Università e della Ricerca (NextGenEU 153D23001490006 M4C2.1.1, NextGenEU I53D23000820006 M4C2.1.1, NextGenEU I53D23001490006 M4C2.1.1, SOE2024\_0000023); Japan: Japan Society for the Promotion of Science (JSPS KAKENHI JP22H01227, JSPS KAKENHI JP22H04944, JSPS KAKENHI JP22KK0227, JSPS KAKENHI JP23KK0245, JSPS KAKENHI JP24K23939); Norway: Research Council of Norway (RCN-314472); Poland: Ministry of Science and Higher Education (IDUB AGH, POB8, D4 no 9722), Polish National Science Centre (NCN 2021/42/E/ST2/00350, NCN OPUS 2023/51/B/ST2/02507, NCN OPUS nr 2022/47/B/ST2/03059, NCN UMO-2019/34/E/ST2/00393, UMO-2022/47/O/ST2/00148, UMO-2023/49/B/ST2/04085, UMO-2023/51/B/ST2/00920, UMO-2024/53/N/ST2/00869); Portugal: Foundation for Science and Technology (FCT); Spain: Agència de Gestió d¿Ajuts Universitaris i de Recerca. (AGAUR - 2023 BP 00141), Generalitat Valenciana (ASFAE/2022/008), Ministry of Science and Innovation (MCIN \& NextGenEU PCI2022-135018-2, MICIN \& FEDER PID2021-125273NB, RYC2019-028510-I, RYC2020-030254-I, RYC2021-031273-I, RYC2022-038164-I), Ministerio de Ciencia, Innovación y Universidades/Agencia Estatal de Investigaci\'on (PID2022-142604OB-C22); Sweden: Carl Trygger Foundation (Carl Trygger Foundation CTS 22:2312), Swedish Research Council (Swedish Research Council 2023-04654, VR 2021-03651, VR 2022-03845, VR 2022-04683, VR 2023-03403, VR 2024-05451), Knut and Alice Wallenberg Foundation (KAW 2018.0458, KAW 2022.0358, KAW 2023.0366); Switzerland: Swiss National Science Foundation (SNSF - PCEFP2\_194658); United Kingdom: Leverhulme Trust (Leverhulme Trust RPG-2020-004), Royal Society (NIF-R1-231091); United States of America: U.S. Department of Energy (ECA DE-AC02-76SF00515), Neubauer Family Foundation.

%
%


%
%
%
%
%
%

%
%

%
%
%
\clearpage
\printbibliography

@article{PHENIX-NPDF1,
  title         = {Centrality Dependence of ${\ensuremath{\pi}}^{0}$ and $\ensuremath{\eta}$ Production at Large Transverse Momentum in $\sqrt{{s}_{NN}}=200\text{ }\text{ }\mathrm{GeV}$ $d+\mathrm{Au}$ Collisions},
  author        = {{PHENIX Collaboration}},
  collaboration = {PHENIX Collaboration},
  journal       = {Phys. Rev. Lett.},
  volume        = {98},
  issue         = {17},
  pages         = {172302},
  numpages      = {7},
  year          = {2007},
  month         = {Apr},
  publisher     = {American Physical Society},
  doi           = {10.1103/PhysRevLett.98.172302},
  eprint        = {nucl-ex/0610036},
  archiveprefix = {arXiv}
}

@article{PHENIX-NPDDF2,
  title         = {Spectra and ratios of identified particles in Au+Au and $d$+Au collisions at $\sqrt{{s}_{NN}}=200$ GeV},
  author        = {{PHENIX Collaboration}},
  collaboration = {PHENIX Collaboration},
  journal       = {Phys. Rev. C},
  volume        = {88},
  issue         = {2},
  pages         = {024906},
  numpages      = {16},
  year          = {2013},
  month         = {Aug},
  publisher     = {American Physical Society},
  doi           = {10.1103/PhysRevC.88.024906},
  eprint        = {1304.3410},
  archiveprefix = {arXiv},
  primaryclass  = {nucl-ex}
}

@article{STAR-NPDF1,
  title         = {Inclusive ${\ensuremath{\pi}}^{0}$, $\ensuremath{\eta}$, and direct photon production at high transverse momentum in $p+p$ and $d+\mathrm{Au}$ collisions at $\sqrt{{s}_{\mathit{NN}}}=200$ GeV},
  author        = {{STAR Collaboration}},
  collaboration = {STAR Collaboration},
  journal       = {Phys. Rev. C},
  volume        = {81},
  issue         = {6},
  pages         = {064904},
  numpages      = {26},
  year          = {2010},
  month         = {Jun},
  publisher     = {American Physical Society},
  doi           = {10.1103/PhysRevC.81.064904},
  eprint        = {0912.3838},
  archiveprefix = {arXiv},
  primaryclass  = {hep-ex}
}

@article{STAR-NPDF2,
  title         = {Identified hadron spectra at large transverse momentum in p+p and d+Au collisions at sNN=200 GeV},
  journal       = {Physics Letters B},
  volume        = {637},
  number        = {3},
  pages         = {161-169},
  year          = {2006},
  issn          = {0370-2693},
  doi           = {https://doi.org/10.1016/j.physletb.2006.04.032},
  collaboration = {STAR Collaboration},
  author        = {{STAR Collaboration}},
  eprint        = {nucl-ex/0601033},
  archiveprefix = {arXiv}
}

@Article{Evans:2008zzb,
      author         = "Evans, Lyndon and Bryant, Philip",
      title          = "{LHC Machine}",
      journal        = "JINST",
      volume         = "3",
      pages          = "S08001",
      doi            = "10.1088/1748-0221/3/08/S08001",
      year           = "2008",
      SLACcitation   = "%%CITATION = JINST,3,S08001;%%",
}

@Article{Cacciari:2008gp,
     author    = "Cacciari, Matteo and Salam, Gavin P. and Soyez, Gregory",
     title     = "{The anti-\(k_{t}\) jet clustering algorithm}",
     journal   = "JHEP",
     volume    = "04",
     year      = "2008",
     pages     = "063",
     eprint    = "0802.1189",
     archivePrefix = "arXiv",
     primaryClass  =  "hep-ph",
     doi       = "10.1088/1126-6708/2008/04/063",
     SLACcitation  = "%%CITATION = 0802.1189;%%"
}

@Article{Fastjet,
      author         = "Cacciari, Matteo and Salam, Gavin P. and Soyez, Gregory",
      title          = "{FastJet user manual}",
      journal        = "Eur. Phys. J. C",
      volume         = "72",
      year           = "2012",
      pages          = "1896",
      doi            = "10.1140/epjc/s10052-012-1896-2",
      eprint         = "1111.6097",
      archivePrefix  = "arXiv",
      primaryClass   = "hep-ph",
      reportNumber   = "CERN-PH-TH-2011-297",
      SLACcitation   = "%%CITATION = ARXIV:1111.6097;%%"
}

@article{Ball:2012cx,
      author         = "{NNPDF Collaboration} and Ball, Richard D. and others",
      title          = "{Parton distributions with LHC data}",
      journal        = "Nucl. Phys. B",
      volume         = "867",
      year           = "2013",
      pages          = "244",
      doi            = "10.1016/j.nuclphysb.2012.10.003",
      eprint         = "1207.1303",
      archivePrefix  = "arXiv",
      primaryClass   = "hep-ph",
      reportNumber   = "EDINBURGH-2012-08, IFUM-FT-997, FR-PHENO-2012-014,
                        RWTH-TTK-12-25, CERN-PH-TH-2012-037, SFB-CPP-12-47\,
                        --CERN-PH-TH-2012-037",
      SLACcitation   = "%%CITATION = ARXIV:1207.1303;%%"
}

@article{Sjostrand:2014zea,
      author         = "Sj{\"o}strand, Torbj{\"o}rn and Ask, Stefan and Christiansen,
                        Jesper R. and Corke, Richard and Desai, Nishita and Ilten,
                        Philip and Mrenna, Stephen and Prestel, Stefan and
                        Rasmussen, Christine O. and Skands, Peter Z.",
      title          = "{An introduction to PYTHIA 8.2}",
      journal        = "Comput. Phys. Commun.",
      volume         = "191",
      year           = "2015",
      pages          = "159",
      doi            = "10.1016/j.cpc.2015.01.024",
      eprint         = "1410.3012",
      archivePrefix  = "arXiv",
      primaryClass   = "hep-ph",
      reportNumber   = "LU-TP-14-36, MCNET-14-22, CERN-PH-TH-2014-190,
                        FERMILAB-PUB-14-316-CD, DESY-14-178, SLAC-PUB-16122,
                        --FERMILAB-PUB-14-316-CD",
      SLACcitation   = "%%CITATION = ARXIV:1410.3012;%%"
}

@article{Agostinelli:2002hh,
      author         = "Agostinelli, S. and others",
      title          = "{\textsc{Geant4} -- a simulation toolkit}",
      journal        = "Nucl. Instrum. Meth. A",
      volume         = "506",
      year           = "2003",
      pages          = "250",
      doi            = "10.1016/S0168-9002(03)01368-8",
      reportNumber   = "SLAC-PUB-9350, FERMILAB-PUB-03-339",
      SLACcitation   = "%%CITATION = NUIMA,A506,250;%%"
}

@Inproceedings{Adye:2011gm,
    author = "Adye, Tim",
    title = "{Unfolding algorithms and tests using RooUnfold}",
    booktitle = "{Proceedings, 2011 Workshop on Statistical Issues Related to Discovery Claims in Search Experiments and Unfolding (PHYSTAT 2011)}",
    venue = "CERN, Geneva, Switzerland",
    eventdate = "2011-01-17/2011-01-20",
    pages = "313-318",
    eprint = "1105.1160",
    archivePrefix = "arXiv",
    primaryClass = "physics.data-an",
    doi = "10.5170/CERN-2011-006.313",
}

@article{DAgostini:1994fjx,
      author         = "D'Agostini, G.",
      title          = "{A multidimensional unfolding method based on Bayes' theorem}",
      journal        = "Nucl. Instrum. Meth. A",
      volume         = "362",
      year           = "1995",
      pages          = "487-498",
      doi            = "10.1016/0168-9002(95)00274-X",
      reportNumber   = "DESY-94-099",
      SLACcitation   = "%%CITATION = NUIMA,A362,487;%%"
}

@Article{PERF-2007-01,
    author         = "{ATLAS Collaboration}",
    title          = "{The ATLAS Experiment at the CERN Large Hadron Collider}",
    journal        = "JINST",
    volume         = "3",
    year           = "2008",
    pages          = "S08003",
    doi            = "10.1088/1748-0221/3/08/S08003",
    primaryClass   = "hep-ex",
}

@Article{HION-2012-13,
    author         = "{ATLAS Collaboration}",
    title          = "{Observation of Associated Near-Side and Away-Side Long-Range Correlations in \(\sqrt{s_{\text{NN}}} = 5.02\,\text{TeV}\) Proton--Lead Collisions with the ATLAS Detector}",
    journal        = "Phys. Rev. Lett.",
    volume         = "110",
    year           = "2013",
    pages          = "182302",
    doi            = "10.1103/PhysRevLett.110.182302",
    reportNumber   = "CERN-PH-EP-2012-366",
    eprint         = "1212.5198",
    archivePrefix  = "arXiv",
    primaryClass   = "hep-ex",
}

@Article{HION-2013-08,
    author         = "{ATLAS Collaboration}",
    title          = "{Centrality and rapidity dependence of inclusive jet production in \(\sqrt{s_\text{NN}} = 5.02\,\text{TeV}\) proton--lead collisions with the ATLAS detector}",
    journal        = "Phys. Lett. B",
    volume         = "748",
    year           = "2015",
    pages          = "392",
    doi            = "10.1016/j.physletb.2015.07.023",
    reportNumber   = "CERN-PH-EP-2014-275",
    eprint         = "1412.4092",
    archivePrefix  = "arXiv",
    primaryClass   = "hep-ex",
}

@Article{HION-2013-09,
    author         = "{ATLAS Collaboration}",
    title          = "{\(Z\) boson production in \(p+Pb\) collisions at \(\sqrt{s_{\text{NN}}}=5.02\,\text{TeV}\) measured with the ATLAS detector}",
    journal        = "Phys. Rev. C",
    volume         = "92",
    year           = "2015",
    pages          = "044915",
    doi            = "10.1103/PhysRevC.92.044915",
    reportNumber   = "CERN-PH-EP-2015-146",
    eprint         = "1507.06232",
    archivePrefix  = "arXiv",
    primaryClass   = "hep-ex",
}

@Article{PERF-2014-07,
    author         = "{ATLAS Collaboration}",
    title          = "{Topological cell clustering in the ATLAS calorimeters and its performance in LHC Run~1}",
    journal        = "Eur. Phys. J. C",
    volume         = "77",
    year           = "2017",
    pages          = "490",
    doi            = "10.1140/epjc/s10052-017-5004-5",
    reportNumber   = "CERN-PH-EP-2015-304",
    eprint         = "1603.02934",
    archivePrefix  = "arXiv",
    primaryClass   = "hep-ex",
}

@Article{HION-2015-02,
    author         = "{ATLAS Collaboration}",
    title          = "{Measurement of the dependence of transverse energy production at large pseudorapidity on the hard-scattering kinematics of proton--proton collisions at \(\sqrt{s} = 2.76\,\text{TeV}\) with ATLAS}",
    journal        = "Phys. Lett. B",
    volume         = "756",
    year           = "2016",
    pages          = "10",
    doi            = "10.1016/j.physletb.2016.02.056",
    reportNumber   = "CERN-PH-EP-2015-293",
    eprint         = "1512.00197",
    archivePrefix  = "arXiv",
    primaryClass   = "hep-ex",
}

@Article{TRIG-2016-01,
    author         = "{ATLAS Collaboration}",
    title          = "{Performance of the ATLAS trigger system in 2015}",
    journal        = "Eur. Phys. J. C",
    volume         = "77",
    year           = "2017",
    pages          = "317",
    doi            = "10.1140/epjc/s10052-017-4852-3",
    reportNumber   = "CERN-EP-2016-241",
    eprint         = "1611.09661",
    archivePrefix  = "arXiv",
    primaryClass   = "hep-ex",
}

@Article{HION-2017-06,
    author         = "{ATLAS Collaboration}",
    title          = "{Dijet azimuthal correlations and conditional yields in \(pp\) and \(p\) + Pb collisions at \(\sqrt{s_{\text{NN}}} = 5.02\,\text{TeV}\) with the ATLAS detector}",
    journal        = "Phys. Rev. C",
    volume         = "100",
    year           = "2019",
    pages          = "034903",
    doi            = "10.1103/PhysRevC.100.034903",
    reportNumber   = "CERN-EP-2018-324",
    eprint         = "1901.10440",
    archivePrefix  = "arXiv",
    primaryClass   = "nucl-ex",
}

@Article{HION-2018-27,
    author         = "{ATLAS Collaboration}",
    title          = "{Two-particle azimuthal correlations in photonuclear ultraperipheral Pb+Pb collisions at \(5.02\,\text{TeV}\) with ATLAS}",
    journal        = "Phys. Rev. C",
    volume         = "104",
    year           = "2021",
    pages          = "014903",
    doi            = "10.1103/PhysRevC.104.014903",
    reportNumber   = "CERN-EP-2020-246",
    eprint         = "2101.10771",
    archivePrefix  = "arXiv",
    primaryClass   = "nucl-ex",
}

@Article{PIX-2018-001,
    author         = "Abbott, B. and others",
    title          = "{Production and integration of the ATLAS Insertable B-Layer}",
    journal        = "JINST",
    volume         = "13",
    year           = "2018",
    pages          = "T05008",
    doi            = "10.1088/1748-0221/13/05/T05008",
    eprint         = "1803.00844",
    archivePrefix  = "arXiv",
    primaryClass   = "physics.ins-det",
}

@Article{HION-2022-14,
    author         = "{ATLAS Collaboration}",
    title          = "{Comparison of inclusive and photon-tagged jet suppression in \(5.02\,\text{TeV}\) Pb+Pb collisions with ATLAS}",
    journal        = "Phys. Lett. B",
    volume         = "846",
    year           = "2023",
    pages          = "138154",
    doi            = "10.1016/j.physletb.2023.138154",
    reportNumber   = "CERN-EP-2023-029",
    eprint         = "2303.10090",
    archivePrefix  = "arXiv",
    primaryClass   = "nucl-ex",
}

@Article{TOPQ-2023-32,
    author         = "{ATLAS Collaboration}",
    title          = "{Observation of \(t\bar{t}\) production in the lepton+jets and dilepton channels in \(p\)+Pb collisions at \(\sqrt{s_{\text{NN}}} = 8.16\,\text{TeV}\) with the ATLAS detector}",
    journal        = "JHEP",
    volume         = "11",
    year           = "2024",
    pages          = "101",
    doi            = "10.1007/JHEP11(2024)101",
    reportNumber   = "CERN-EP-2024-097",
    eprint         = "2405.05078",
    archivePrefix  = "arXiv",
    primaryClass   = "nucl-ex",
}

@Article{HION-2024-11,
    author         = "{ATLAS Collaboration}",
    title          = "{Characterization of nuclear breakup as a function of hard-scattering kinematics using dijets measured by ATLAS in \(p\)+Pb collisions}",
    year           = "2025",
    reportNumber   = "CERN-EP-2025-077",
    eprint         = "2504.02638",
    archivePrefix  = "arXiv",
    primaryClass   = "nucl-ex",
}

@Booklet{ATL-SOFT-PUB-2026-001,
    author         = "{ATLAS Collaboration}",
    title          = "{ATLAS Computing Acknowledgements}",
    howpublished   = "{ATL-SOFT-PUB-2026-001}",
    url            = "https://cds.cern.ch/record/2952666",
    year           = "2026",
}

@Report{ATLAS-TDR-19,
    author         = "{ATLAS Collaboration}",
    title          = "{ATLAS Insertable B-Layer: Technical Design Report}",
    type           = "ATLAS-TDR-19; CERN-LHCC-2010-013",
    year           = "2010",
    url            = "https://cds.cern.ch/record/1291633",
    related        = "ATLAS-TDR-19-addm",
    relatedstring  = "Addendum:",
}

@Article{CMS-HIN-13-001,
    author         = "{CMS Collaboration}",
    title          = "{Studies of dijet transverse momentum balance and pseudorapidity distributions in \(p\)Pb collisions at \(\sqrt{s_{\text{NN}}} = 5.02\,\text{TeV}\)}",
    journal        = "Eur. Phys. J. C",
    volume         = "74",
    year           = "2014",
    pages          = "2951",
    doi            = "10.1140/epjc/s10052-014-2951-y",
    reportNumber   = "CERN-PH-EP-2013-236",
    eprint         = "1401.4433",
    archivePrefix  = "arXiv",
    primaryClass   = "hep-ex",
}

@Article{CMS-HIN-13-007,
    author         = "{CMS Collaboration}",
    title          = "{Study of \(W\) boson production in \(p\)Pb collisions at \(\sqrt{s_{\text{NN}}}\) \(5.02\,\text{TeV}\)}",
    journal        = "Phys. Lett. B",
    volume         = "750",
    year           = "2015",
    pages          = "565",
    doi            = "10.1016/j.physletb.2015.09.057",
    reportNumber   = "CERN-PH-EP-2015-054",
    eprint         = "1503.05825",
    archivePrefix  = "arXiv",
    primaryClass   = "hep-ex",
}

@Article{CMS-HIN-14-001,
    author         = "{CMS Collaboration}",
    title          = "{Measurement of inclusive jet production and nuclear modifications in \(p\)Pb collisions at \(\sqrt{s_{\text{NN}}} = 5.02\,\text{TeV}\)}",
    journal        = "Eur. Phys. J. C",
    volume         = "76",
    year           = "2016",
    pages          = "372",
    doi            = "10.1140/epjc/s10052-016-4205-7",
    reportNumber   = "CERN-PH-EP-2015-334",
    eprint         = "1601.02001",
    archivePrefix  = "arXiv",
    primaryClass   = "hep-ex",
}

@Article{CMS-HIN-17-002,
    author         = "{CMS Collaboration}",
    title          = "{Observation of Top Quark Production in Proton--Nucleus Collisions}",
    journal        = "Phys. Rev. Lett.",
    volume         = "119",
    year           = "2017",
    pages          = "242001",
    doi            = "10.1103/PhysRevLett.119.242001",
    reportNumber   = "CERN-EP-2017-239",
    eprint         = "1709.07411",
    archivePrefix  = "arXiv",
    primaryClass   = "hep-ex",
}

@Booklet{ATL-PHYS-PUB-2014-021,
    author         = "{ATLAS Collaboration}",
    title          = "{ATLAS Pythia~8 tunes to \(7\,\text{TeV}\) data}",
    howpublished   = "{ATL-PHYS-PUB-2014-021}",
    url            = "https://cds.cern.ch/record/1966419",
    year           = "2014",
}

@Booklet{ATL-PHYS-PUB-2021-011,
    author         = "{ATLAS Collaboration}",
    title          = "{Evaluating statistical uncertainties and correlations using the bootstrap method}",
    howpublished   = "{ATL-PHYS-PUB-2021-011}",
    url            = "https://cds.cern.ch/record/2759945",
    year           = "2021",
}

@article{Alvioli:2025ggv,
  archiveprefix = {arXiv},
  author        = {Alvioli, M. and Guzey, V. and Strikman, M.},
  doi           = {10.1103/jxyf-8xvm},
  eprint        = {2504.07514},
  journal       = {Phys. Rev. C},
  number        = {3},
  pages         = {034901},
  primaryclass  = {hep-ph},
  title         = {{Probing fluctuating protons using forward neutrons in soft and hard inelastic proton-nucleus scattering}},
  volume        = {112},
  year          = {2025}
}

@article{Nagle:2018nvi,
  archiveprefix = {arXiv},
  author        = {Nagle, James L. and Zajc, William A.},
  doi           = {10.1146/annurev-nucl-101916-123209},
  eprint        = {1801.03477},
  journal       = {Ann. Rev. Nucl. Part. Sci.},
  pages         = {211--235},
  primaryclass  = {nucl-ex},
  title         = {{Small System Collectivity in Relativistic Hadronic and Nuclear Collisions}},
  volume        = {68},
  year          = {2018}
}

@article{ATLAS:2015hkr,
  archiveprefix = {arXiv},
  author        = {{ATLAS Collaboration}},
  collaboration = {ATLAS},
  doi           = {10.1140/epjc/s10052-016-4002-3},
  eprint        = {1508.00848},
  journal       = {Eur. Phys. J. C},
  number        = {4},
  pages         = {199},
  primaryclass  = {hep-ex},
  reportnumber  = {CERN-PH-EP-2015-160},
  title         = {{Measurement of the centrality dependence of the charged-particle pseudorapidity distribution in proton\textendash{}lead collisions at \(\sqrt{s_\text{NN}} = 5.02\,\text{TeV}\) with the ATLAS detector}},
  volume        = {76},
  year          = {2016}
}

@article{PHENIX:2015fgy,
  archiveprefix = {arXiv},
  author        = {{PHENIX Collaboration}},
  collaboration = {PHENIX},
  doi           = {10.1103/PhysRevLett.116.122301},
  eprint        = {1509.04657},
  journal       = {Phys. Rev. Lett.},
  number        = {12},
  pages         = {122301},
  primaryclass  = {nucl-ex},
  title         = {{Centrality-dependent modification of jet-production rates in deuteron-gold collisions at \(\sqrt{s_\text{NN}} = 200\,\text{GeV}\)}},
  volume        = {116},
  year          = {2016}
}

@article{ALICE:2014xsp,
  archiveprefix = {arXiv},
  author        = {{ALICE Collaboration}},
  collaboration = {ALICE},
  doi           = {10.1103/PhysRevC.91.064905},
  eprint        = {1412.6828},
  journal       = {Phys. Rev. C},
  number        = {6},
  pages         = {064905},
  primaryclass  = {nucl-ex},
  reportnumber  = {CERN-PH-EP-2014-281},
  title         = {{Centrality dependence of particle production in p-Pb collisions at \(\sqrt{s_\text{NN}} = 5.02\,\text{TeV}\)}},
  volume        = {91},
  year          = {2015}
}

@article{ATLAS:2022iyq,
  archiveprefix = {arXiv},
  author        = {{ATLAS Collaboration}},
  doi           = {10.1103/PhysRevLett.131.072301},
  eprint        = {2206.01138},
  journal       = {Phys. Rev. Lett.},
  pages         = {072301},
  primaryclass  = {nucl-ex},
  reportnumber  = {CERN-EP-2022-086},
  title         = {{Strong constraints on jet quenching in centrality-dependent \(p\)+Pb collisions at \(5.02\,\text{TeV}\) from ATLAS}},
  volume        = {131},
  year          = {2023}
}

@article{ATLAS:2019vcm,
  archiveprefix = {arXiv},
  author        = {{ATLAS Collaboration}},
  collaboration = {ATLAS},
  doi           = {10.1140/epjc/s10052-020-7624-4},
  eprint        = {1910.13978},
  journal       = {Eur. Phys. J. C},
  number        = {1},
  pages         = {73},
  primaryclass  = {nucl-ex},
  reportnumber  = {CERN-EP-2019-217},
  title         = {{Transverse momentum and process dependent azimuthal anisotropies in \(\sqrt{s_\text{NN}} = 8.16\,\text{TeV}\) $p$+Pb collisions with the ATLAS detector}},
  volume        = {80},
  year          = {2020}
}

@article{Klasen:2023uqj,
  archiveprefix = {arXiv},
  author        = {Klasen, M. and Paukkunen, H.},
  doi           = {10.1146/annurev-nucl-102122-022747},
  eprint        = {2311.00450},
  journal       = {Ann. Rev. Nucl. Part. Sci.},
  pages         = {49--87},
  primaryclass  = {hep-ph},
  reportnumber  = {MS-TP-23-45},
  title         = {{Nuclear Parton Distribution Functions After the First Decade of LHC Data}},
  volume        = {74},
  year          = {2024}
}

@article{ALICE:2017svf,
  archiveprefix = {arXiv},
  author        = {{ALICE Collaboration}},
  collaboration = {ALICE},
  doi           = {10.1016/j.physletb.2018.05.059},
  eprint        = {1712.05603},
  journal       = {Phys. Lett. B},
  pages         = {95--113},
  primaryclass  = {nucl-ex},
  reportnumber  = {CERN-EP-2017-324},
  title         = {{Constraints on jet quenching in p-Pb collisions at $\mathbf{\sqrt{s_{NN}}}$ = 5.02 TeV measured by the event-activity dependence of semi-inclusive hadron-jet distributions}},
  volume        = {783},
  year          = {2018}
}

@article{STAR:2024nwm,
  archiveprefix = {arXiv},
  author        = {{STAR Collaboration}},
  collaboration = {STAR},
  doi           = {10.1103/PhysRevC.110.044908},
  eprint        = {2404.08784},
  journal       = {Phys. Rev. C},
  number        = {4},
  pages         = {044908},
  primaryclass  = {nucl-ex},
  title         = {{Correlations of event activity with hard and soft processes in p+Au collisions at sNN=200 GeV at the RHIC STAR experiment}},
  volume        = {110},
  year          = {2024}
}

@article{Brodsky:2022bum,
  archiveprefix = {arXiv},
  author        = {Brodsky, Stanley J. and de Teramond, Guy F.},
  doi           = {10.3390/physics4020042},
  eprint        = {2202.13283},
  journal       = {MDPI Physics},
  number        = {2},
  pages         = {633--646},
  primaryclass  = {hep-ph},
  reportnumber  = {SLAC-PUB-17656},
  title         = {{Onset of Color Transparency in Holographic Light-Front QCD}},
  volume        = {4},
  year          = {2022}
}

@article{ATLAS:2014cpa,
  archiveprefix = {arXiv},
  author        = {{ATLAS Collaboration}},
  collaboration = {ATLAS},
  doi           = {10.1016/j.physletb.2015.07.023},
  eprint        = {1412.4092},
  journal       = {Phys. Lett. B},
  pages         = {392--413},
  primaryclass  = {hep-ex},
  reportnumber  = {CERN-PH-EP-2014-275},
  title         = {{Centrality and rapidity dependence of inclusive jet production in \(\sqrt{s_\text{NN}} = 5.02\,\text{TeV}\) proton-lead collisions with the ATLAS detector}},
  volume        = {748},
  year          = {2015}
}

@article{ALICE:2021poe,
  archiveprefix = {arXiv},
  author        = {{ALICE Collaboration}},
  doi           = {10.1007/JHEP08(2022)086},
  eprint        = {2107.10757},
  journal       = {JHEP},
  pages         = {086},
  primaryclass  = {nucl-ex},
  reportnumber  = {CERN-EP-2021-144},
  title         = {{Study of very forward energy and its correlation with particle production at midrapidity in pp and p-Pb collisions at the LHC}},
  volume        = {08},
  year          = {2022}
}

@article{Zheng:2014cha,
  archiveprefix = {arXiv},
  author        = {Zheng, L. and Aschenauer, E. C. and Lee, J. H.},
  doi           = {10.1140/epja/i2014-14189-3},
  eprint        = {1407.8055},
  journal       = {Eur. Phys. J. A},
  number        = {12},
  pages         = {189},
  primaryclass  = {hep-ex},
  title         = {{Determination of electron-nucleus collision geometry with forward neutrons}},
  volume        = {50},
  year          = {2014}
}

@article{Alvioli:2024cmd,
  archiveprefix = {arXiv},
  author        = {Alvioli, M. and Guzey, V. and Strikman, M.},
  doi           = {10.1103/PhysRevC.110.025205},
  eprint        = {2402.19060},
  journal       = {Phys. Rev. C},
  number        = {2},
  pages         = {025205},
  primaryclass  = {hep-ph},
  title         = {{Slicing Pomerons in ultraperipheral collisions using forward neutrons from nuclear breakup}},
  volume        = {110},
  year          = {2024}
}

@article{Albacete:2017qng,
  archiveprefix = {arXiv},
  author        = {Albacete, Javier L. and others},
  doi           = {10.1016/j.nuclphysa.2017.11.015},
  eprint        = {1707.09973},
  journal       = {Nucl. Phys. A},
  pages         = {18--85},
  primaryclass  = {hep-ph},
  title         = {{Predictions for cold nuclear matter effects in $p+$Pb Collisions at  \(\sqrt{s_\text{NN}} = 8.16\,\text{TeV}\)}},
  volume        = {972},
  year          = {2018}
}

@article{Alvioli:2017wou,
  archiveprefix = {arXiv},
  author        = {Alvioli, Massimiliano and Frankfurt, Leonid and Perepelitsa, Dennis V and Strikman, Mark},
  doi           = {10.1103/PhysRevD.98.071502},
  eprint        = {1709.04993},
  journal       = {Phys. Rev. D},
  number        = {7},
  pages         = {071502},
  primaryclass  = {hep-ph},
  title         = {{Global analysis of color fluctuation effects in proton\textendash{} and deuteron\textendash{}nucleus collisions at RHIC and the LHC}},
  volume        = {98},
  year          = {2018}
}

@report{Adriani:2135330,
  author = {{LHCf Collaboration}},
  title  = {{Study of forward physics in $\sqrt{s_\text{NN}}=8.1$ TeV
            proton-Lead ion collisions with the LHCf detector at the
            LHC}},
  type   = {CERN-LHCC-2016-003, LHCC-I-027},
  url    = {https://cds.cern.ch/record/2135330},
  year   = {2016}
}

@article{Ethier:2020way,
  archiveprefix = {arXiv},
  author        = {Ethier, Jacob J. and Nocera, Emanuele R.},
  doi           = {10.1146/annurev-nucl-011720-042725},
  eprint        = {2001.07722},
  journal       = {Ann. Rev. Nucl. Part. Sci.},
  pages         = {43--76},
  primaryclass  = {hep-ph},
  reportnumber  = {Nikhef/2020-003},
  title         = {{Parton Distributions in Nucleons and Nuclei}},
  volume        = {70},
  year          = {2020}
}

@article{LHCb:2017yua,
  archiveprefix = {arXiv},
  author        = {{LHCb Collaboration}},
  collaboration = {LHCb},
  doi           = {10.1007/JHEP10(2017)090},
  eprint        = {1707.02750},
  journal       = {JHEP},
  pages         = {090},
  primaryclass  = {hep-ex},
  reportnumber  = {LHCB-PAPER-2017-015, LHCb-PAPER-2017-015, CERN-EP-2017-135},
  title         = {{Study of prompt D$^{0}$ meson production in $p$Pb collisions at $ \sqrt{s_{\mathrm{NN}}}=5 $ TeV}},
  volume        = {10},
  year          = {2017}
}

@article{Perepelitsa:2024eik,
  archiveprefix = {arXiv},
  author        = {Perepelitsa, Dennis V.},
  doi           = {10.1103/PhysRevC.110.L011901},
  eprint        = {2404.17660},
  journal       = {Phys. Rev. C},
  number        = {1},
  pages         = {L011901},
  primaryclass  = {nucl-th},
  title         = {{Contribution to differential $\pi^0$ and $\gamma^\mathrm{dir}$ modification in small systems from color fluctuation effects}},
  volume        = {110},
  year          = {2024}
}

@article{ATLAS:2023zfx,
  archiveprefix = {arXiv},
  author        = {{ATLAS Collaboration}},
  doi           = {10.1103/PhysRevLett.132.102301},
  eprint        = {2309.00033},
  journal       = {Phys. Rev. Lett.},
  number        = {10},
  pages         = {102301},
  primaryclass  = {nucl-ex},
  reportnumber  = {CERN-EP-2023-178},
  title         = {{Measurement of the centrality dependence of the dijet yield in p+Pb collisions at $\sqrt{s_{\text{NN}}}=8.16$ TeV with the ATLAS Detector}},
  volume        = {132},
  year          = {2024}
}

@inproceedings{Mueller:1982,
  author    = {Mueller, Alfred H.},
  booktitle = {{17th Rencontres de Moriond on Elementary Particle Physics: I. Electroweak Interactions and Grand Unified Theories}},
  pages     = {13--43},
  title     = {{Topics in High-Energy Perturbative QCD Including Interactions with Nuclear Matter}},
  year      = {1982}
}

@inproceedings{Brodsky:1982,
  author       = {Brodsky, Stanley J.},
  booktitle    = {{XIII International Symposium on Multiparticle Dynamics}},
  month        = {9},
  reportnumber = {SLAC-PUB-2970},
  title        = {{Testing Quantum Chromodynamics}},
  year         = {1982}
}

@article{PhysRevLett.47.297,
  author    = {Bertsch, G. and Brodsky, S. J. and Goldhaber, A. S. and Gunion, J. G.},
  doi       = {10.1103/PhysRevLett.47.297},
  journal   = {Phys. Rev. Lett.},
  month     = {8},
  numpages  = {0},
  pages     = {297--300},
  publisher = {American Physical Society},
  title     = {Diffractive Excitation in Quantum Chromodynamics},
  volume    = {47},
  year      = {1981}
}

@article{PhysRevC.107.064902,
  archiveprefix = {arXiv},
  author        = {{ALICE Collaboration}},
  collaboration = {ALICE Collaboration},
  doi           = {10.1103/PhysRevC.107.064902},
  eprint        = {2209.04250},
  issue         = {6},
  journal       = {Phys. Rev. C},
  month         = {Jun},
  numpages      = {18},
  pages         = {064902},
  primaryclass  = {nucl-ex},
  publisher     = {American Physical Society},
  title         = {Neutron emission in ultraperipheral Pb-Pb collisions at $\sqrt{{s}_{NN}}=5.02$ TeV},
  volume        = {107},
  year          = {2023}
}

@article{HION-2022-15-Published,
  archiveprefix = {arXiv},
  author        = {{ATLAS Collaboration}},
  doi           = {10.1103/PhysRevD.111.052006},
  eprint        = {2409.11060},
  journal       = {Phys. Rev. D.},
  pages         = {052006},
  primaryclass  = {nucl-ex},
  reportnumber  = {CERN-EP-2024-221},
  title         = {{Measurement of photonuclear jet production in ultra-peripheral Pb+Pb collisions at \(\sqrt{s_{\text{NN}}} = 5.02\,\text{TeV}\) with the ATLAS detector}},
  volume        = {111},
  year          = {2025}
}

@article{SOFT-2022-02-Published,
  archiveprefix = {arXiv},
  author        = {{ATLAS Collaboration}},
  doi           = {10.1140/epjc/s10052-024-13701-w},
  eprint        = {2404.06335},
  journal       = {Eur. Phys. J. C},
  pages         = {234},
  primaryclass  = {hep-ex},
  reportnumber  = {CERN-EP-2024-100},
  title         = {{Software and computing for Run~3 of the ATLAS experiment at the LHC}},
  volume        = {85},
  year          = {2025}
}

@article{Cortese_2019,
  author    = {Cortese, Pietro and for the ALICE Collaboration},
  doi       = {10.1088/1742-6596/1162/1/012006},
  journal   = {Journal of Physics: Conference Series},
  month     = {jan},
  number    = {1},
  pages     = {012006},
  publisher = {IOP Publishing},
  title     = {Performance of the ALICE Zero Degree Calorimeters and upgrade strategy},
  url       = {https://dx.doi.org/10.1088/1742-6596/1162/1/012006},
  volume    = {1162},
  year      = {2019}
}

@article{PhysRevLett.107.172301,
  archiveprefix = {arXiv},
  author        = {{PHENIX Collaboration}},
  doi           = {10.1103/PhysRevLett.107.172301},
  eprint        = {1105.5112},
  issue         = {17},
  journal       = {Phys. Rev. Lett.},
  month         = {Oct},
  numpages      = {7},
  pages         = {172301},
  primaryclass  = {nucl-ex},
  publisher     = {American Physical Society},
  title         = {Suppression of Back-to-Back Hadron Pairs at Forward Rapidity in $d+\mathrm{Au}$ Collisions at $\sqrt{{s}_{NN}}=200\text{ }\text{ }\mathrm{GeV}$},
  volume        = {107},
  year          = {2011}
}

@article{PhysRevLett.123.102001,
  archiveprefix = {arXiv},
  author        = {Ridder, A. Gehrmann-De and Gehrmann, T. and Glover, E. W. N. and Huss, A. and Pires, J.},
  doi           = {10.1103/PhysRevLett.123.102001},
  eprint        = {1905.09047},
  issue         = {10},
  journal       = {Phys. Rev. Lett.},
  month         = {Sep},
  numpages      = {5},
  pages         = {102001},
  primaryclass  = {hep-ph},
  publisher     = {American Physical Society},
  title         = {Triple Differential Dijet Cross Section at the LHC},
  volume        = {123},
  year          = {2019}
}

\clearpage
 
\begin{flushleft}
\hypersetup{urlcolor=black}
{\Large The ATLAS Collaboration}

\bigskip

\AtlasOrcid[0000-0002-6665-4934]{G.~Aad}$^\textrm{\scriptsize 104}$,
\AtlasOrcid[0000-0001-7616-1554]{E.~Aakvaag}$^\textrm{\scriptsize 17}$,
\AtlasOrcid[0000-0002-5888-2734]{B.~Abbott}$^\textrm{\scriptsize 123}$,
\AtlasOrcid[0000-0002-0287-5869]{S.~Abdelhameed}$^\textrm{\scriptsize 119a}$,
\AtlasOrcid[0000-0002-1002-1652]{K.~Abeling}$^\textrm{\scriptsize 55}$,
\AtlasOrcid[0000-0001-5763-2760]{N.J.~Abicht}$^\textrm{\scriptsize 49}$,
\AtlasOrcid[0000-0002-8496-9294]{S.H.~Abidi}$^\textrm{\scriptsize 30}$,
\AtlasOrcid[0009-0003-6578-220X]{M.~Aboelela}$^\textrm{\scriptsize 45}$,
\AtlasOrcid[0000-0002-9987-2292]{A.~Aboulhorma}$^\textrm{\scriptsize 36e}$,
\AtlasOrcid[0000-0001-5329-6640]{H.~Abramowicz}$^\textrm{\scriptsize 157}$
% ...
\bigskip
\\

% ...
$^{17}$Department for Physics and Technology, University of Bergen, Bergen; Norway.\\
% ...
$^{30}$Physics Department, Brookhaven National Laboratory, Upton NY; United States of America.\\
% ...
$^{36}$$^{(a)}$Facult\'e des Sciences Ain Chock, Universit\'e Hassan II de Casablanca;$^{(b)}$Facult\'{e} des Sciences, Universit\'{e} Ibn-Tofail, K\'{e}nitra;$^{(c)}$Facult\'e des Sciences Semlalia, Universit\'e Cadi Ayyad, LPHEA-Marrakech;$^{(d)}$LPMR, Facult\'e des Sciences, Universit\'e Mohamed Premier, Oujda;$^{(e)}$Facult\'e des sciences, Universit\'e Mohammed V, Rabat;$^{(f)}$Institute of Applied Physics, Mohammed VI Polytechnic University, Ben Guerir; Morocco.\\
% ...
$^{45}$Physics Department, Southern Methodist University, Dallas TX; United States of America.\\
% ...
$^{49}$Fakult\"{a}t Physik, Technische Universit{\"a}t Dortmund, Dortmund; Germany.\\
% ...
$^{55}$II. Physikalisches Institut, Georg-August-Universit\"{a}t G\"ottingen, G\"ottingen; Germany.\\
% ...
$^{104}$CPPM, Aix-Marseille Universit\'e, CNRS/IN2P3, Marseille; France.\\
% ...
$^{119}$$^{(a)}$New York University Abu Dhabi, Abu Dhabi;$^{(b)}$United Arab Emirates University, Al Ain; United Arab Emirates.\\
% ...
$^{123}$Homer L. Dodge Department of Physics and Astronomy, University of Oklahoma, Norman OK; United States of America.\\
% ...
$^{157}$Raymond and Beverly Sackler School of Physics and Astronomy, Tel Aviv University, Tel Aviv; Israel.\\
% ...
$^{a}$ Also at Affiliated with an institute formerly covered by a cooperation agreement with CERN.\\
$^{b}$ Also at An-Najah National University, Nablus; Palestine.\\
$^{c}$ Also at Borough of Manhattan Community College, City University of New York, New York NY; United States of America.\\
$^{d}$ Also at Center for Interdisciplinary Research and Innovation (CIRI-AUTH), Thessaloniki; Greece.\\
$^{e}$ Also at Centre of Physics of the Universities of Minho and Porto (CF-UM-UP); Portugal.\\
$^{f}$ Also at CERN, Geneva; Switzerland.\\
% ...

\end{flushleft}


%

%
%

%

%
%
%
%

\end{document}